\documentclass[twocolumn]{aastex631}

\newcommand{\magra}{\texttt{MAGRATHEA}}

\graphicspath{{../Figures/}}
\accepted{April 22, 2025}

\shorttitle{Uncertainties in internal structure for T1-f}
\shortauthors{Rice et al.}

\begin{document}

\title{Uncertainties in the Inference of Internal Structure: The Case of TRAPPIST-1 f}

\correspondingauthor{David R. Rice}
\email{davidr@post.openu.ac.il}

\author[0000-0001-6009-8685]{David R. Rice}
\affiliation{Astrophysics Research Center (ARCO), Department of Natural Sciences, The Open University of Israel, Raanana 4353701, Israel}

\author[0000-0001-9446-6853]{Chenliang Huang}
\affiliation{Shanghai Astronomical Observatory, Chinese Academy of Sciences, Shanghai 200030, People’s Republic of China}

\author[0000-0003-2202-3847]{Jason H. Steffen}
\affiliation{Department of Physics and Astronomy, University of Nevada, Las Vegas, 4505 South Maryland Parkway, Las Vegas, NV 89154, USA}
\affiliation{Nevada Center for Astrophysics, University of Nevada, Las Vegas, 4505 South Maryland Parkway, Las Vegas, NV 89154, USA}

\author[0000-0001-9504-3174]{Allona Vazan}
\affiliation{Astrophysics Research Center (ARCO), Department of Natural Sciences, The Open University of Israel, Raanana 4353701, Israel}



\begin{abstract}

We use the TRAPPIST-1 system as a model observation of Earth-like planets. The densities of these planets being 1-10\% less than the Earth suggest that the outer planets may host significant hydrospheres. We explore the uncertainty in water mass fraction from observed mass and radius. We investigate the interior structure of TRAPPIST-1 f using the open-source solver \magra\ and varying assumptions in the interior model. We find that TRAPPIST-1 f likely has a water mass fraction of 16.2\% ± 9.9\% when considering all possible core mass fractions and requires 6.9\% ± 2.0\% water at an Earth-like mantle to core ratio. We quantify uncertainties from observational precision, model assumptions, and experimental and theoretical data on the bulk modulus of planet building materials. We show that observational uncertainties are smaller than model assumptions of mantle mineralogy and core composition but larger than hydrosphere, temperature, and equation of state assumptions/uncertainties. Our findings show that while precise mass and radius measurements are crucial, uncertainties in planetary models can often outweigh those from observations, emphasizing the importance of refining both theoretical models and experimental data to better understand exoplanet interiors.

\end{abstract}



\section{Introduction}

Although there are now over 5,800 confirmed planets, there are fewer than 150 small planets of less than 10 M$_\oplus$ with both an accurate mass and radius (\textless20\% uncertainties)\footnote{NASA Exoplanet Archive, DOI 10.26133/NEA12, as of 05 February 2025}. This number will grow rapidly over the next decade as more extreme precision radial velocity spectrographs see first light and complete follow-up surveys of transiting exoplanets \citep{Rauer2024, Howard2025}. The densities of small planets (\textless10 M$_\oplus$) suggest that terrestrial worlds can span a wide range of compositions, from Mercury-like to Ganymede-like \citep{Barros2022,Piaulet2023}.

When observational studies determine or refine the masses of small planets, they will often include a section describing the planets' possible compositions.  In a small sample of observational studies there are a number of compositional analyses based solely on comparisons to mass-radius models: \citet{Winters2022} and \citet{Serrano2022} use models of \citet{Zeng2016}; \citet{Chaturvedi2022}, \citet{Beard2022}, and \citet{Kemmer2022} use the updated \citet{Zeng2019}; and \citet{Luque2022} uses \citet{Turbet2020}.  While other studies use interior models to give further constraints on possible compositions: \citet{Barros2022} uses an improved model based on \citet{Dorn2017}; \citet{Cadieux2022} uses \citet{Plotnykov2020} with an added water layer; \citet{Essack2023} use their own model; and \citet{Desai2024} uses \citet{Magrathea}.

Underlying the interior models used in the above studies are multiple computational techniques (e.g. shooting methods and relaxation methods), numerous experimental measurements and theoretical estimates of the equations of state for planet-building materials, and differing treatments of temperature. In \citet{Magrathea}, we describe \magra, an open-source interior structure solver that can be customized to user-defined planet models. Our code features adaptable phase diagrams for the core, mantle, hydrosphere, and atmosphere and a large library of equations of state for planetary materials. With four distinct layers, the code is most applicable to planets roughly from sub-Mars to sub-Neptune ($\sim$0.1-15 M$_\oplus$) although future work aims to expand this range.

Previous studies of small planets (\textless10 M$_\oplus$) focused on how model degeneracies impact the mass-radius relationship for specific planet types. \citet{Grasset2009} explores the mass–radius relationship for silicate-, iron-, and water-worlds across a wide range of planetary mass with various temperatures. \citet{Howe2014} compares mass-radius relationships for various mantle equations of state and models four-layer sub-Neptunes with a hydrogen-helium envelope. \citet{Thomas2016} focuses on water-rich exoplanets with the addition of a water vapor atmosphere and illustrates how small changes in temperature and hydrosphere assumptions can produce large uncertainties in mass–radius estimates. Further investigations on mass-radius relationships include \citet{Unterborn2018, Hakim2018, Noack2020, Aguichine2021, Shah2021}.

In this work, we use \magra's flexible and computationally cheap model to carry through observational, model, and experimental uncertainties to inferences of a single planet's composition with both mass and radius constraints. Accurately representing the uncertainty in a planet's composition is important to carry through to the study of mantle convection, atmosphere evolution, and habitability.

To quantify uncertainties, we turn to the TRAPPIST-1 (T1) system as a model observation. \citet{Gillon2016} first reports the transits of the inner three planets observed with the TRAnsiting Planets and PlanestIsimals Small Telescope. \citet{trappist1first} then measures the masses and radii of all seven currently known planets with transits observed by the Spitzer Space Telescope. \citet{TRAPPIST1} reanalyzes the transit-timing variations confirming that the seven planets ranged from just 0.33 to 1.56 M$_\oplus$. T1 is a nearby M8 star which has a radius only 16\% larger than Jupiter. The transit depth of the T1 planets is comparable to a Saturn around the Sun. In addition, the outermost planet orbits in just 18.8 days leading to many transits and strong perturbations in transit times from the closely packed nature of the system. These characteristics make the T1 system a valuable benchmark for studying Earth-sized exoplanets with well-measured bulk densities.

The literature on the T1 system is extensive and growing, and we list the briefest review of relevant literature here. With such a small, cool star, several of the T1 planets are in the ``habitable zone'' and could have the right temperature for liquid water or ice Ih on the surface. The planets are bombarded with UV radiation from their active M-star \citep{Vida2018}, but if the planets formed with an abundance of water, they could hold on to significant hydrospheres through present day \citep{Bolmont2017, Dong2018, Childs2023}. A primordial, hydrogen-dominated, cloud-free atmosphere is ruled out for all planets \citep{deWit2018}. In the near future, detection of CO$_2$, H$_2$O and CH$_4$ is feasible with JWST for T1-e and -f \citep{Fauchez2019, Mikal-Evans2022, Lustig-Yaeger2019}. 

T1-f is situated in most estimates of the habitable zone with a null-albedo surface temperature of 218 K requiring only a small atmosphere for the surface to be warm enough for liquid water. \citet{Agol2021} measures T1-f's mass to be near 1 M$_\oplus$, radius near 1 R$_\oplus$, but a bulk density 9\% lower than Earth. With its larger size (compared to T1-d and -e) and being in the middle of the system, T1-f has the lowest observational uncertainties with 3.0\% uncertainty in mass and 1.2\% uncertainty in radius. 

T1-f presents an ideal observation at the edge of our current observing capabilities. We use T1-f as a benchmark in order to compare and quantify the effect of model parameters on the inferences of a planet composition without considerations of formation, stellar compositions, and evolution. This study also seeks to elucidate the precision needed by future observations to understand planet composition and the limitations from laboratory studies of material properties.

Our paper is laid out as follows. In Section \ref{sec:model}, we describe our interior model and composition finder. In Section \ref{sec:interiors}, we find the likely interior structures of the T1 planets with our default model and detail our findings for T1-f. Then we examine our uncertainties and their effect on the inferred water mass fraction (wmf)---observational uncertainty in Section \ref{sec:obs}, model and compositional uncertainty in Section \ref{sec:modelunc}, and experimental uncertainty on the bulk modulus in the mantle and core in Section \ref{sec:eos}. We discuss and conclude in Section \ref{sec:conclusion}. Additionally, in Appendix \ref{app:plots}, we discuss standards for data representation and describe how our representations agree with and differ from previous literature. 

\clearpage
\section{Default Model}
\label{sec:model}

We use the one dimensional interior structure model \magra\ developed in \citet{Magrathea}\footnote{\magra\ can be accessed at \url{https://github.com/Huang-CL/Magrathea}}. The solver takes as input the mass in each differentiated layer of the planet and temperature discontinuities at layer boundaries. The solver uses a shooting to a fitting point method to solve the equations of hydrostatic equilibrium in order to find the planet's radius and internal conditions at steps of enclosed mass. The model uses an adiabatic temperature profile for each layer by default. The phase of the material in each layer is determined by the pressure and temperature at each step.

\magra\ hosts a library of equation of state (EoS) fitted formulas for planet building materials. Here we describe the default parameters and materials used in this work. The innermost differentiated layer is the core. In the core, we use pure iron in the hexagonal close-packed phase (hcp-Fe) from \citet{Smith2018}. At temperatures above the melting curve, we use liquid iron from \citet{Dorogokupets17}.  Above the core in the mantle, we use magnesium silicates. In the upper mantle (updated since \citet{Magrathea}), we use forsterite (Fo), wadsleyite (Wds), and ringwoodite (Rgd)---polymorphs of Mg$_2$SiO$_4$. We use the material properties and phase transitions from \citet{Dorogokupets2015}. At pressures above approximately 20-25 GPa, the stable mineral is MgSiO$_3$ in the bridgmanite (Brg) phase---material parameters from \citet{Oganov04} transition from \citet{Dorogokupets2015}. At pressures above 110-150 GPa, the stable phase is post-perovskite magnesium silicate (PPv) with material parameters from \citet{Sakai16} and transition from \citet{Ono05}.

The outermost layer in this study is the hydrosphere, a layer of pure H$_2$O. We use phase transitions from \citet{Dunaeva2010} for water and low-pressure ice and from \citet{Grande2022} for high-pressure ice. We use EoS for water from \citet{Valencia2007}, for Ice Ih from \citet{Feistel2006}, for Ice VI and VII from \citet{Bezacier2014}, and for Ice X from \citet{Grande2022}. For Ice X, we do not include thermodynamic properties as measurements have large uncertainties and treat temperature as constant in this relatively small layer occurring only when the wmf is above 20\%.

\begin{figure}[ht!]
\plotone{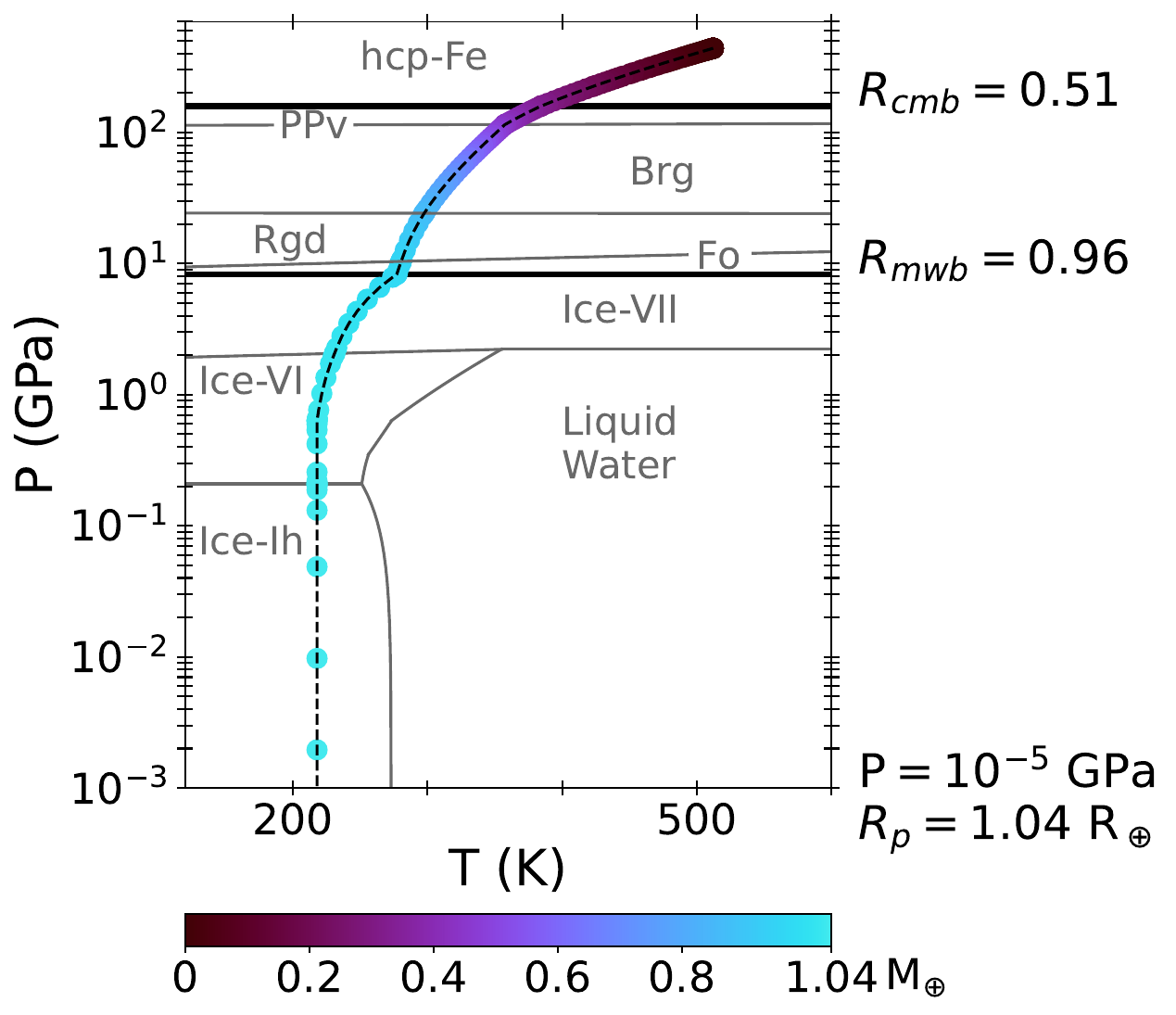}
\caption{Interior conditions of a 1.039 M$_\oplus$ planet with 6.9\% of mass in the hydrosphere, 62.4\% in mantle, and 30.7\% in core. Shown by the colored points is the pressure and temperature at each step of enclosed mass found by \magra. The radius of the core-mantle boundary (R$_{cmb}$), mantle-hydrosphere boundary (R$_{mwb}$), and radius at 100 mbar (R$_p$) are noted. The phase used in regions of P-T for each layer is shown. This planet does not reach pressures in the hydrosphere to form Ice X and the small layer of wadsleyite between forsterite and ringwoodite is omitted. 
}
\label{fig:t1fphases}
\end{figure}

The default model is visualized in Fig.~\ref{fig:t1fphases} where we show the output of \magra\ for a 1.039 M$_\oplus$ planet with 6.9\% of mass in the hydrosphere, 62.4\% in mantle, and 30.7\% in core. This planet represents the median solution of T1-f with an Earth-like core to mantle ratio described later in Section \ref{sec:t1fdefault}. The planet passes through a number of the phases described above as pressure and temperature increase deeper in the planet (with decreased enclosed mass).

Further explanations of EoS parameters are in \citet{Magrathea}. While many aspects of the ``default'' model described above will stay the same throughout this work changes to this model will be described in each section.

\subsection{Composition Finder}\label{sec:finder}

There are multiple solutions of the interior structure of a three-layer planet for a given mass and radius \citep{Seager2007}. We developed a composition finder routine for \magra\ which, given the mass and radius of a planet, finds an array of solutions to the interior structure\footnote{The routine can be used with the input file \textit{mode4.cfg} in \magra}. 

The composition finder ingests a tab separated file of mass and radius samples. These samples can be generated by fitting a Gaussian or other distribution to the observed median and standard deviation, or samples can be generated directly from an observational pipeline. In this work, we draw mass and radius measurements directly from the posterior samples used in and provided by \citet{Agol2021}. Using posterior samples more realistically captures the distribution of observed mass and radius. Besides simulating the skewness of each value, a combined analysis of mass and radius will have a correlation between mass and radius. This correlation often is not accounted for in interior studies.



For a given planet mass, we define the refractory core mass fraction (r-cmf) as the dependent variable across which we find the water mass fraction (wmf) to match a given planet radius. We use the term ``refractory'' here to reference that in our definition r-cmf is the core mass fraction of the refractory components of the planet or the planet without a hydrosphere and atmosphere. We define r-cmf as
\begin{equation}
    \label{eq:RCMF}
    \textrm{r-cmf}=\frac{M_{core}}{M_{mantle}+M_{core}}
\end{equation}
or the denominator can equally be expressed as the total mass minus the mass of the hydrosphere (and atmosphere if considering one), $M_{tot}-M_{hydro}$. 

We sample the r-cmf space by uniformly stepping from 1 to 0 in steps of either 0.01 or 0.02 leading to 101 or 51 solutions of wmf for a mass and radius input. The solver starts the planet with 100\% of the input mass in the core and solves a planet with the given temperatures using the full \magra\ solver. This approach gives the smallest radius for a given mass of planet. If the target radius is smaller than a 100\% core planet, the mass and radius measurements are non-physical for the given planet model. In this case, an error is returned and we move on to the next mass and radius.

If the target radius is larger than the 100\% core planet then 0.1\% of the mass is converted to hydrosphere mass, and we find the radius of this new planet with r-cmf still equal to unity. We then use a secant method \citep{secant} to find the optimal wmf to match the target radius ($R_{targ}$) with the remaining mass in the core. From the previous two radii ($R_i$, $R_j$) for the water mass fractions ($f_{w,i}$, $f_{w,j}$), the next water mass fraction ($f_{w,k}$) is found from
\begin{equation}
    \label{eq:secant}
    f_{w,k}=f_{w,j}-(f_{w,j}-f_{w,i})\frac{(R_j-R_{targ})}{(R_j-R_i)}.
\end{equation}
This is repeated until the planet radius compared to $R_{targ}$ is within a specified tolerance. For a tolerance of 0.1\%, the finder typically converges within 5-6 iterations. 

The finder is then swept across decreasing r-cmf. The finder begins each step with no volatile layer. In the second iteration onward after determining the wmf, $f_{w}$, the core mass fraction, $f_c$, is given by
\begin{equation}
    \label{eq:coresec}
    f_{c}=\textrm{r-cmf}(1-f_{w}),
\end{equation}
and the mantle mass fraction, $f_m$, by
\begin{equation}
    \label{eq:mantsec}
    f_{m}=(1-\textrm{r-cmf})(1-f_{w}).
\end{equation}
The optimal wmf is recorded for each r-cmf. When the initial planet's radius with no hydrosphere is larger than the target radius, there are no longer solutions at lower r-cmf and the finder moves to the next mass and radius measurement. 

The routine is able to be done in parallel with each mass and radius sample run on a different thread. We accomplish multithreading with \textsc{OpenMP} \citep{openmp}. With average run times of 1 second and approximately 600 runs needed to find solutions at steps of 0.01 in r-cmf for a mass and radius measurement, the finder takes approximately 10 minutes per mass and radius. In this work we run 1000-5000 samples of mass and radius which takes a maximum of 3 days across 12 cores. This scales linearly with step size and with number of samples. 

Previous works which use interior solvers to characterize planets from both mass and radius include \citet{Dorn2015, Santos2015, Dorn2017b, Dorn2017, Brugger2017, Plotnykov2020, Baumeister2020} along with many individual observational studies. Many of these studies use a Markov chain Monte Carlo (MCMC) method to characterize planets. In the MCMC method, a greater number of model parameters can be explored than mass fraction in each layer. In \citet{Dorn2017}, they use MCMC to constrain the mass fraction of water and atmosphere along with the mantle Fe/Si and Mg/Si ratios. While we generate solutions uniformly across r-cmf, treating it as entirely unknown, the MCMC treatment can use a prior distribution such as one constrained by stellar abundances. However, current codes that run this treatment require 10-100 times more computation time than our composition finder. 

In addition to MCMC methods, machine learning methods are being increasingly used to more quickly characterize a planet's composition. Mixture density networks (MDNs) have been developed as efficient surrogates for MCMC, capable of reproducing posterior distributions at a fraction of the computational cost \citep{Baumeister2023,2023ApJS..269....1Z}. Unlike MCMC, which relies on stochastic sampling to explore parameter space, MDNs are initially computationally expensive requiring a large training set of planetary interior models but then learn how to quickly map observables---such as mass, radius, and elemental abundance ratios---to a probabilistic distribution of interior structure parameters.

Our composition finder is computational cheap while finding all three layer interior solutions with a given resolution of r-cmf. This scheme can also be used in other applications as which layer is unknown (see Section \ref{sec:atm}) and which layers are to be held in a constant ratio can be set within the code. How our composition finder relates to our data representation throughout the work is described in Appendix \ref{app:plots}.

\section{Interiors of the TRAPPIST-1 Planets}
\label{sec:interiors}

While T1-f is the focus of this study, we first aim to place T1-f into context of the other six planets in the system. We use the default model and composition finder to find the likely three-layer interiors for the five outer planets T1-d, -e, -f, -g, and -h. For the inner two planets, we include a 2-layer model of only mantle and core. The equilibrium temperature of the inner two planets of the T1 system are near the the boiling point of water, and thus any hydrosphere would form a steam atmosphere. In the case of a steam atmosphere, the radius of the planet depends more on the temperature profile of the gas than the wmf \citep{Aguichine2021}. However \citet{greene2023thermal}, uses JWST to observe the thermal emission from secondary eclipses of T1-b. For all seven planets, we draw 5000 mass and radius measurements from the 10,000 posterior samples directly generated by \citet{Agol2021}\footnote{Mass and radius posterior samples accessed from \url{https://github.com/ericagol/TRAPPIST1_Spitzer}}. The properties of the seven planets and our findings are summarized in Table \ref{tab:t1}.

\begin{table*}
	\centering
	\caption{Planet mass, radius, density, correlation coefficient of mass and radius, and null albedo equilibrium temperature of the T1 planets as found by \citet{Agol2021}. The core mass fraction (cmf) assuming zero water mass fraction (wmf) is reported for all seven planets. The overall likely wmf and the wmf at Earth-like refractory core mass fraction (r-cmf) reported for the outer five planets. Values represent 50th percentile while uncertainties calculated from the 15.9 and 84.1 percentile. \label{tab:t1}}
	\begin{tabular}{lccccccc} 
	\hline
	Planet: & b & c & d & e & f & g & h \\
        \hline
        R (R$_\oplus$) & 1.116$^{+0.014}_{-0.012}$ & 1.097$^{+0.014}_{-0.012}$ & 0.788$^{+0.011}_{-0.010}$ & 0.920$^{+0.013}_{-0.012}$ & 1.045$^{+0.013}_{-0.012}$ & 1.129$^{+0.015}_{-0.013}$ & 0.755$^{+0.014}_{-0.014}$ \\ 
        M (M$_\oplus$) & 1.374$\pm$0.069 & 1.308$\pm$0.056 & 0.388$\pm$0.012 & 0.692$\pm$0.022 & 1.039$\pm$0.031 & 1.321$\pm$0.038 & 0.326$\pm$0.020 \\ 
        $\rho$ ($\rho_\oplus$) & 0.987$^{+0.048}_{-0.050}$ & 0.991$^{+0.040}_{-0.043}$ & 0.792$^{+0.028}_{-0.030}$ & 0.889$^{+0.030}_{-0.033}$ & 0.911$^{+0.025}_{-0.029}$ & 0.917$^{+0.025}_{-0.029}$ & 0.755$^{+0.049}_{-0.054}$ \\ 
        r$_{\mathrm{m,r}}$ & 0.36 & 0.42 & 0.49 & 0.51 & 0.58 & 0.58 & 0.20 \\ 
        T$_{eq}$ (K) & 398 & 340 & 286 & 250 & 218 & 197 & 172 \\ 
        cmf$_{\text{wmf}=0}$ (\%) & 19.97$^{+5.02}_{-6.98}$ & 20.98$^{+4.99}_{-5.01}$ & 11.00$^{+5.00}_{-5.00}$ & 17.98$^{+4.02}_{-5.00}$ & 13.98$^{+3.99}_{-4.00}$ & 9.99$^{+4.00}_{-4.00}$ & 10.00$^{+8.98}_{-6.00}$ \\ 
        wmf (\%) &  &  & 15.44$^{+9.59}_{-11.29}$ & 14.75$^{+9.28}_{-10.66}$ & 16.23$^{+9.92}_{-11.61}$ & 17.48$^{+10.55}_{-12.38}$ & 16.75$^{+10.16}_{-11.72}$ \\ 
        wmf$_{\text{r-cmf}=0.33}$ (\%) &  &  & 7.17$^{+2.27}_{-1.99}$ & 5.02$^{+2.10}_{-1.76}$ & 6.93$^{+1.98}_{-1.57}$ & 8.84$^{+2.02}_{-1.63}$ & 9.17$^{+4.68}_{-4.04}$ \\ 
        \hline    
	\end{tabular}
\end{table*}

\begin{figure}[ht!]
\plotone{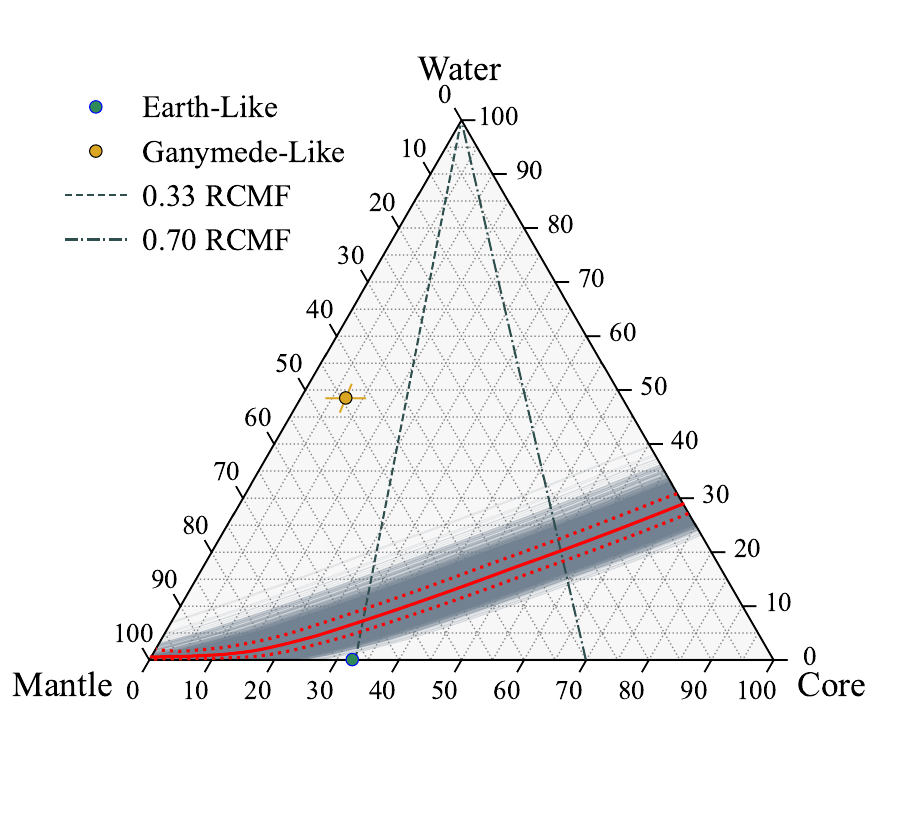}
\plotone{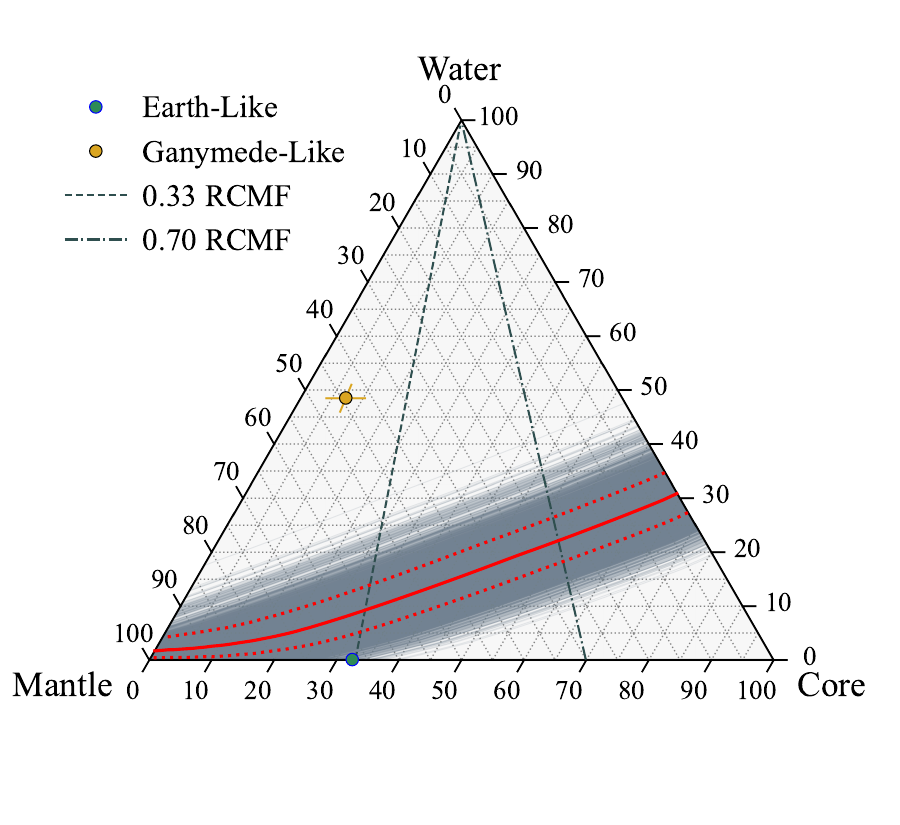}
\caption{\textit{Top}, interior structures which match 5000 draws of mass and radius for T1-d on a water-mantle-core ternary plot (for further description see Appendix \ref{app:plots}). \textit{Bottom}, interior structures for T1-h. \textit{Red lines} show the median (\textit{solid}) and $\pm1\sigma$ bounds (\textit{dotted}) of wmf at integer steps of r-cmf. Earth with no water and Ganymede with 46-51\% wmf are shown along with \textit{gray, dashed} lines of constant Earth-like (0.33) and Mercury-like (0.7) r-cmf.
}
\label{fig:T1dh}
\end{figure}

We show the likely 3-layer interiors for T1-d and T1-h as examples of the five outer planets. Fig.~\ref{fig:T1dh} shows on a water-mantle-core mass ternary diagram the results for T1-d and T1-h. Dashed lines show an Earth-like (0.33) and Mercury-like (0.7) r-cmf and Earth is plotted at a 32.5\% cmf. Ganymede is shown with wmf of 46-51\% and r-cmf of 0.08-0.2 in agreement with models constrained by \textit{Galileo} spacecraft measurements \citep{Schubert2007,Sohl2002,Kuskov2001}. Each gray line is up to 101 solutions to the wmf that match a mass and radius sample at an integer r-cmf. Median and $\pm1\sigma$ bounds when reported throughout this work are determined from the 15.87, 50, 84.13 percentile of samples with wmf solutions at a given r-cmf. We determine the inferred wmf across all r-cmf by taking a weighted percentile. The weight for each wmf solution for a given mass and radius is determined by dividing 101 by the number of solutions for that mass and radius. The weighting ensure that the median mass and radius of inputs is the same as the weighted median mass and radius across all solutions.

Both T1-d and T1-h are a third the mass of Earth. The mass and radius of T1-d is 0.387$\pm$0.012 M$_\oplus$ and 0.788$^{+0.011}_{-0.010}$ R$_\oplus$. We infer wmf across all r-cmf of 15.44$^{+9.59}_{-11.29}$\% for T1-d. T1-h is 4.2\% less massive, 4.7\% less dense, and on average needs 1.31\% more of its mass in water than T1-d. At an Earth-like r-cmf, T1-d requires a 6.80\% wmf and T1-h requires a 8.8\% wmf.


\begin{figure}[ht!]
\plotone{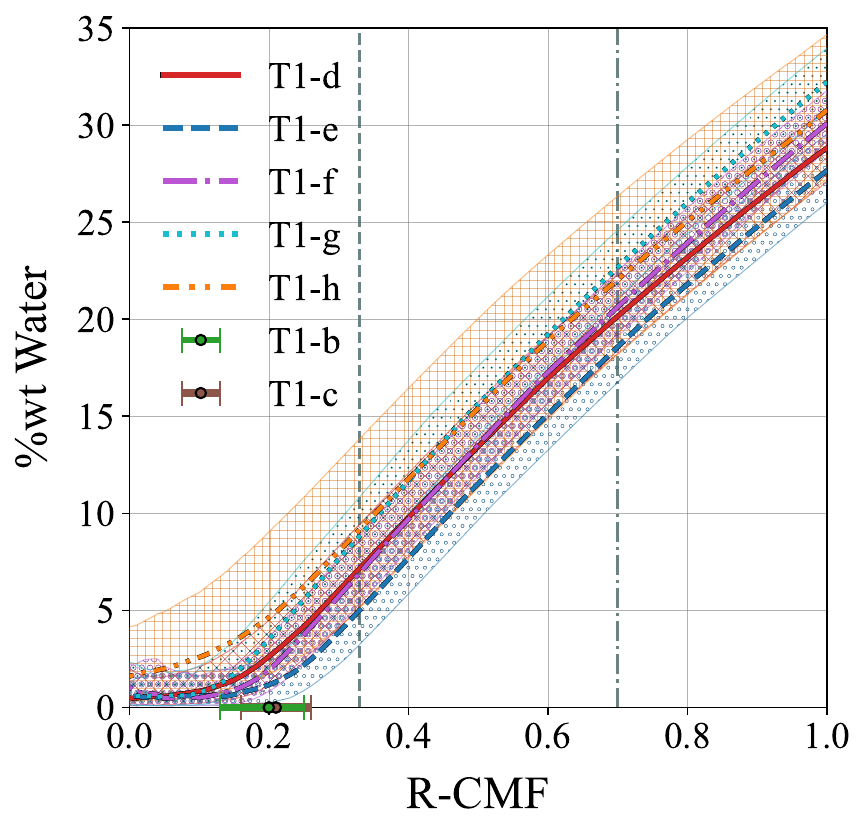}
\caption{The median and 1$\sigma$ wmf of all five outer T1 planets across r-cmf is shown by the various colored lines and hatched regions. T1-b and T1-c's median and 1$\sigma$ cmfs also shown with zero wmf.
}
\label{fig:T1all}
\end{figure}

We show the possible compositions of all seven planets in Fig.~\ref{fig:T1all}. T1-b has a likely cmf of 0.20$^{+0.05}_{-0.07}$ and T1-c has cmf of 0.21$\pm0.05$. The outer planets have lower cmfs if they have no hydrosphere with median cmf of 0.11, 0.18, 0.14, 0.10, and 0.10 from inner to outer. The five planets require 5-10\% wmf at an Earth-like r-cmf. 

The outer planets show a large degree of overlap in potential interior structure. However, if the planets have r-cmf above 0.22, T1-e and -g cannot have the same wmf to 1$\sigma$. The overall likely wmf for the 5 outer planets from inner to outer as reported in Table \ref{tab:t1}: 0.154$^{+0.096}_{-0.113}$, 0.148$^{+0.093}_{-0.107}$, 0.162$^{+0.099}_{-0.116}$, 0.175$^{+0.106}_{-0.124}$, 0.167$^{+0.102}_{-0.117}$. We compare these observationally constrained wmfs to formation scenarios in \citet{Childs2023}.

\subsection{TRAPPIST-1 f} \label{sec:t1fdefault}

We now turn to focus on the likely interior of T1-f being near Earth-like in mass and radius with low uncertainties. T1-f's mass and radius was measured by \citet{TRAPPIST1} to be 0.934$\pm$0.079 M$_\oplus$ and radius of 1.046$\pm$0.03 R$_\oplus$. \citet{Agol2021} shifts the density higher (0.911$^{+0.025}_{-0.029} \rho_\oplus$) with mass of 1.039$\pm$0.031 M$_\oplus$ and radius of 1.044$\pm$0.013 R$_\oplus$. This work cut the uncertainty in both mass and radius in half with additional observations by the \textit{Spitzer Space Telescope}. The overall wmf we infer for T1-f using \citet{Agol2021} is 3.9\% lower than inferred using \citet{TRAPPIST1}, and wmf at an Earth-like r-cmf is nearly half (6.93\% compared to 13.2\%).

\begin{figure}[ht!]
\plotone{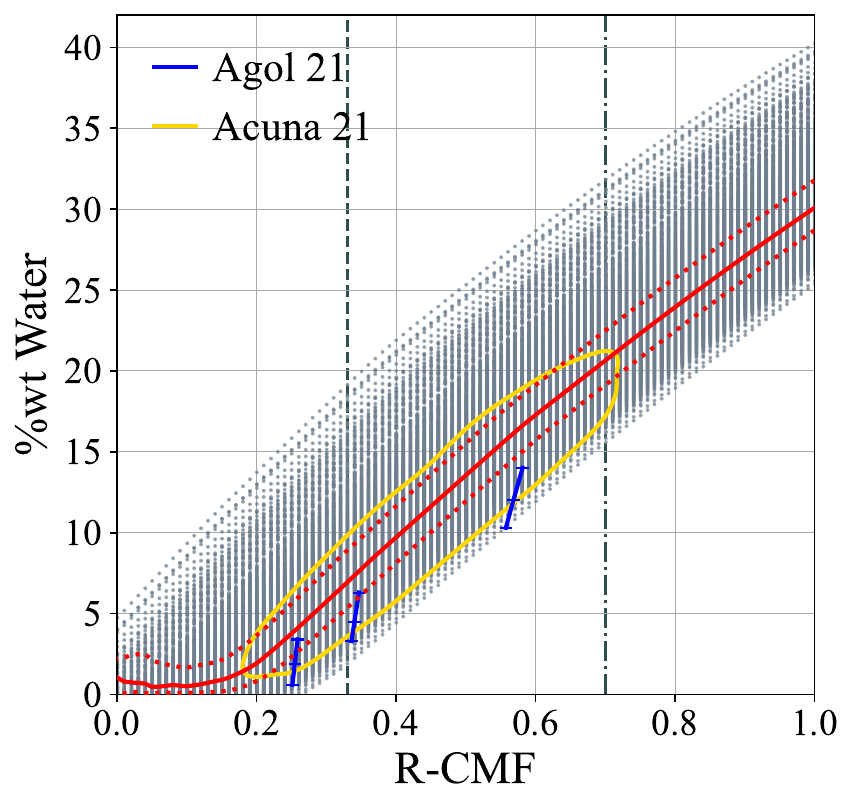}
\caption{The 101 inferred wmfs for T1-f across r-cmf which match each of 5000 draws of mass and radius are shown by gray points. \textit{Red lines} show the median and $\pm1\sigma$ bounds of wmf across uniform r-cmf. Also shown is the wmf found in \citet{Agol2021} at three values of cmf and the $\pm1\sigma$ compositions from \citet{Acuna2021}. Dashed and dash-dotted vertical lines correspond with an Earth-like (0.33) r-cmf and a Mercury-like (0.7) r-cmf.
}
\label{fig:t1fcompare}
\end{figure}

Shown in Fig.~\ref{fig:t1fcompare}, we find that T1-f with our default model, described in Section \ref{sec:model}, and a 218 K surface temperature with no discontinuities in the adiabatic temperature profile has a likely 16.23$^{+9.92}_{-11.61}$\% wmf and requires a 6.93$^{+1.98}_{-1.57}$\% wmf at an Earth-like r-cmf, Fig.~\ref{fig:t1fcompare}. If the planet has no hydrosphere it needs a cmf half that of the Earth. The likely cmf is 14.0$\pm4.0$\% at zero wmf. 1.2\% of mass and radius samples require a volatile layer. 

We show the interior properties of the mediant T1-f solution at an Earth-like r-cmf in Fig.~\ref{fig:t1fphases}. The hydrosphere reaches to a pressure where ice VII forms. The mantle has an upper and lower mantle and reaches pressures where post-perovskite is stable in the lower mantle. The upper mantle is truncated from the presence of the hydrosphere. The central pressure is 438 GPa and central temperature is 512 K.

As shown in Fig.~\ref{fig:t1fcompare}, the wmf that we find across r-cmf matches well with that of \citet{Acuna2021}. They only report $\pm1\sigma$ bounds of three-layer composition and add an upper bound on cmf of 75\% from \citet{Lodders2009}. We expect their underlying cmf values are mostly uniform from 0-75\% in their models without stellar constraints. Their reported values of 0.41$\pm$0.14 for the cmf correspond closely with the 0.38$\pm$0.22 cmf expected from a uniform distribution made slightly higher and tighter by some samples of mass and radius not having solutions below a certain cmf. Our uncertainty in wmf at a given r-cmf is also lower from using the correlated mass radius measurements rather than fitting a Gaussian to the reported mass and radius. 

The wmfs we find are higher than found in \citet{Agol2021} also shown in Fig.~\ref{fig:t1fcompare} and which are plotted as diagonal lines as they report wmf at a given cmf. It is difficult to pinpoint why this disagreement exists without detailed intermodel comparisons. \citet{Acuna2021} suggests the difference is from the EoSs used in the hydrosphere as \citet{Agol2021} uses equations from \citet{Vazan2013} which overestimates the density at pressures higher than 70 GPa, but the hydrosphere does not reach that pressure in most solutions and an over-density would only increase the inferred wmf.  The model in \citet{Agol2021} also uses \citet{Sotin2007} for the mantle which does not include post-perovskite and includes MgO in the mantle. Replacing post-perovskite with bridgmanite leads to only a 0.5\% decrease in the wmf in our models, but the inclusion of the less dense MgO phase mixed with the bridgmanite could decrease the mantle density by 3\% which could be responsible for the lower wmf. However, it is not made clear how much iron is in the mantle in \citet{Agol2021} as considered in \citet{Sotin2007} which would again increase the density of the mantle. Inter-comparisons of interior models should be a focus of exoplanet interior research.



\section{Observational Uncertainties}
\label{sec:obs}

Before quantifying model uncertainties, we show the affect of observational uncertainties on inferred composition. To do so, we first fix the radius uncertainty to 1\%. Radius measurements are often limited by stellar radii measurements. We create samples of mass based on the median reported values of T1-f but change the width, correlation with radius, and skewness of the distribution. We first investigate increasing the mass uncertainty from 3.2\% to 4.2\% and to 10\% of the median value. We draw 2500 samples of mass and radius from Gaussian distributions with the median of T1-f and given uncertainty.  

The inferred wmf results for the Gaussian draws of mass and radius are shown in Table \ref{tab:gauss}. The overall likely wmf is within 0.1\% with approximately the same uncertainty for all samples. However, the uncertainty in wmf at a given r-cmf grows proportionally to the density uncertainty. With 10\% mass uncertainty, 5\% of samples do not have a three-layer solution at a 0.33 r-cmf, thus the wmf becomes skewed. Similarly, the cmf uncertainty stays the same for small mass uncertainties as 97-98\% of samples have a no-water solution. However, only 82\% of samples have a no-water solution for the 10\% mass uncertainty which increases the uncertainty and skew of the measured cmf. 

\begin{figure}[ht!]
\plotone{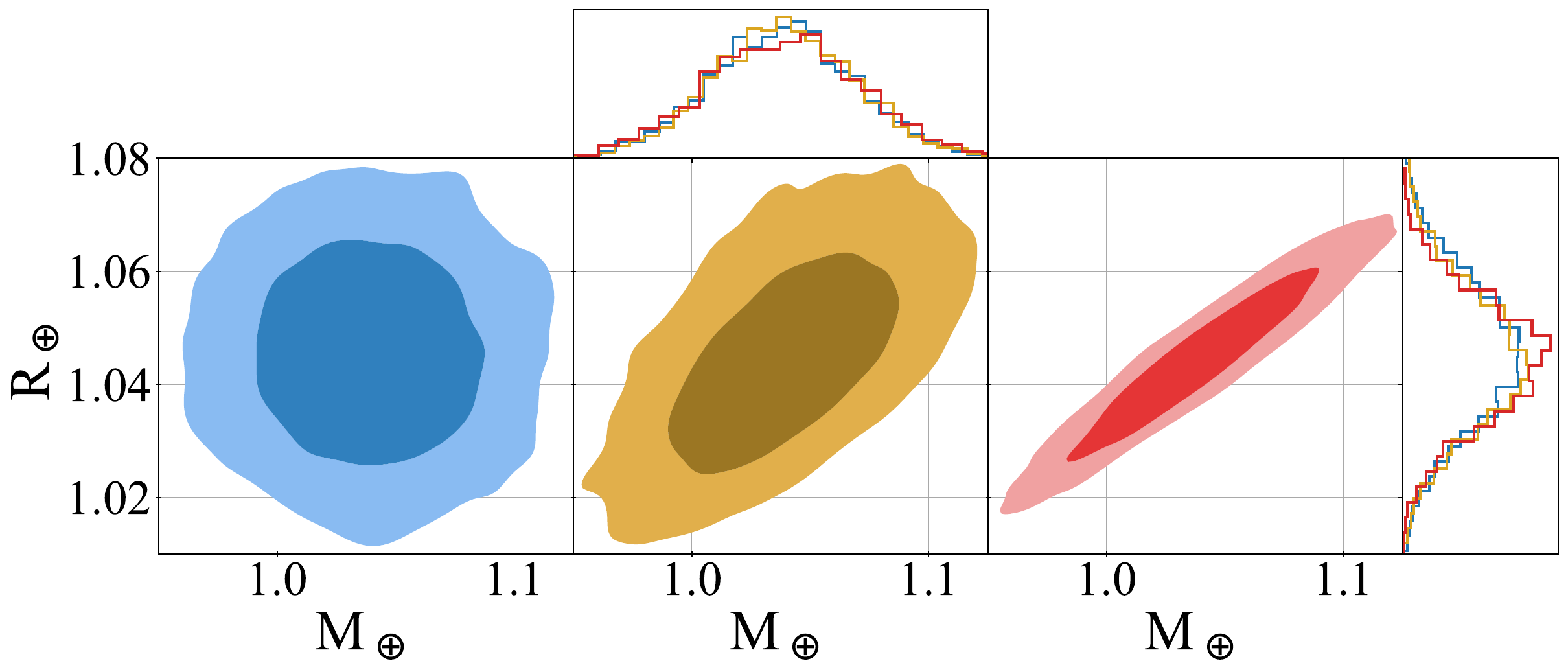}
\plotone{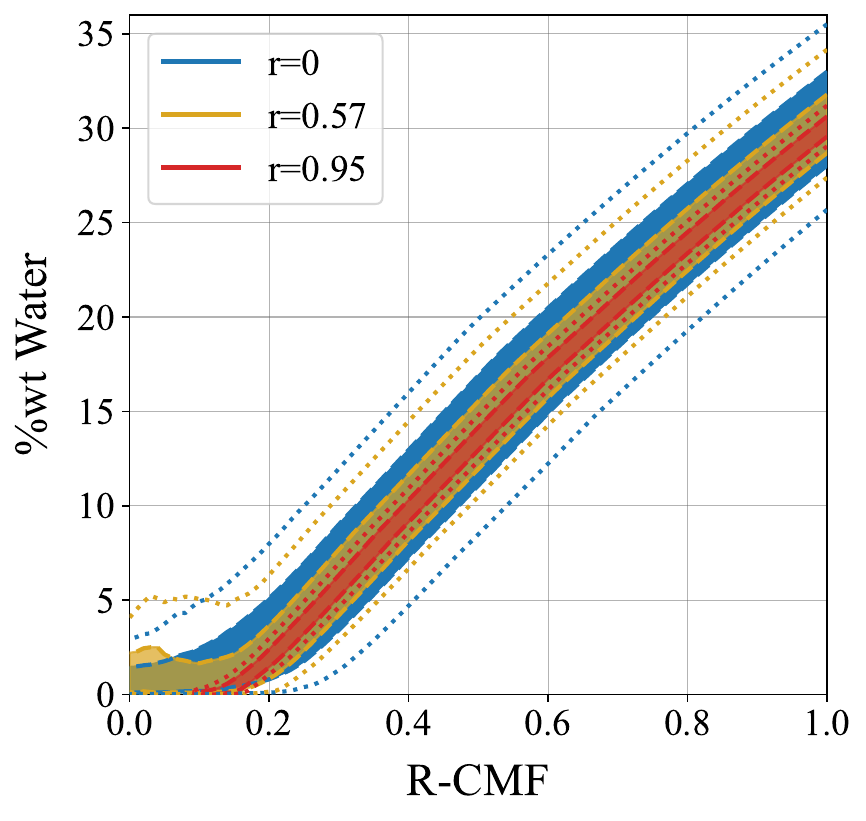}
\caption{\textit{Top}, 1$\sigma$ and 2$\sigma$ ellipses of mass and radius for uncorrelated data (\textit{blue}), \citet{Agol2021} (\textit{gold}), and highly correlated data (\textit{red}). All data have nearly the same median and standard deviation shown in the histograms for both mass and radius. \textit{Bottom}, the $1\sigma$ (\textit{dashed, filled}) and $2\sigma$ (\textit{dotted}) bounds of wmf across r-cmf for the three distributions from \textit{top} with the shown correlation coefficient between mass and radius.
}
\label{fig:t1fcorr}
\end{figure}

The correlation between mass and radius, shown for all seven planets in Table \ref{tab:t1}, leads to a lower wmf uncertainty than the Gaussian samples above. We create a correlated sample of mass and radius for T1-f by constraining the density uncertainty and drawing mass based on a draw of density and radius. We use a density uncertainty of 1\% with radius uncertainty of 1\% which leads to a 3.16\% mass uncertainty in agreement with the expected uncertainty from error propagation. While the Pearson correlation coefficient of mass and radius, r$_{\mathrm{m,r}}$, in \citet{Agol2021} is 0.58, this sample has a correlation coefficient of 0.95. The uncertainty of wmf at a given r-cmf continues to be approximately proportional to the density uncertainty. The density decreases by a factor of 2.9 while the Earth-like wmf uncertainty decreases by a factor of 3.2. See Table \ref{tab:gauss} and Fig.~\ref{fig:t1fcorr} for comparison. 

\begin{table*}
	\centering
    \caption{2,500 samples of mass and radius were generated with the median values of T1-f and the values reported in the table for the uncertainty in mass, radius, and density and the correlation coefficient, r$_{\mathrm{m,r}}$. Uncertainty (Unc \%) in wmf as a percentage of the median value shown at 0.33 (Earth-like) and 0.70 (Mercury-like) r-cmf. Final column is uncertainty in cmf at 0\% wmf. The wmf and cmf uncertainty is the difference between 84.13 percentile and median as a percent of the median. The -1$\sigma$ uncertainty is lower for large uncertainties as wmf below zero are not allowed.}
	\begin{tabular}{cccccccc} 
		\hline
		Desc. & Mass &   Radius & Density &  r$_{\mathrm{m,r}}$   & Earth-wmf & Mercury-wmf  & cmf\\
		&Unc \% & Unc \% & Unc \% &  & Unc \% &  Unc \% &  Unc \% \\
		\hline
         Agol+ 21 &  2.9 & 1.2 & 2.9 & 0.58 & 28.5 & 9.0  &  28.5 \\
	    Gauss 3.2\% &3.2 & 1.0 & 4.5 & 0.00 & 39.5 & 12.4 &  42.8 \\
        Gauss 4.2\% &    4.2 & 1.0 & 5.1 & 0.00 & 43.9 & 13.7 &  42.9 \\
        Gauss 10.3\% &   10.3 & 1.0 & 10.7 & 0.00 & 87.3 &  28.7 &  64.7 \\
        Correlated &   3.2 & 1.0 & 1.0 & 0.95 & 8.7 & 2.8  &  10.7 \\
		\hline
	\end{tabular}
    \label{tab:gauss}
\end{table*}

Lastly, in Fig.~\ref{fig:t1fcompare} there is a slight skewness to the observed data. The Fisher-Pearson coefficient of skewness is 0.48 for the observed radii and 0.01 for observed masses. We create four mock samples of 2500 mass and radii to show the effect of skewness on the results. We draw the mass from a Gaussian with T1-f parameters. We draw the radii from a Gaussian with T1-f parameters, but scale the radii either above or below the mean value. The radii are first normalized, $x=(R-\bar R) /\sigma_R$ and then scaled by $1\pm\alpha x$ for radius above or below the median to achieve either a positive or negative skew. The parameter, $\alpha$, is iterated until the desired Fisher-Pearson coefficient of skewness is reached in the sample.

Three of the samples we create have positive skewness in radii while one has a negative skewness in radii. In our positive skew sample the tail of distribution reaches to larger radii which means the density distribution skews in the opposite direction with lower densities. We run this sample data through our composition finder and report in Table \ref{tab:skew} the wmf with Earth-like (0.33) r-cmf and Mercury-like (0.7) r-cmf and report the upper and lower uncertainty and skewness. The skewness in wmf is similar to that of radius for the positive skews. The 1-$\sigma$ uncertainties increase from under 30\% of the median value in the \citet{Agol2021} data to over 90\% of the median value for highly skewed radii.

The negative skew in radius results in denser planets which are not as sensitive to a changing wmf, which results in less skew in the wmf at a Mercury-like r-cmf. At an Earth-like r-cmf the skew creates samples that are denser than a zero wmf planet, so the distribution is truncated and we find a low skewness. A similar result happens for the positive skew samples when we measure the cmf with no water. These samples have planets that cannot be solved without a hydrosphere and thus create a truncated distribution of cmf. While the skewless sample has 96\% of samples with a non-volatile cmf solution, the highly skewed distribution only has 82\% of samples with a solution.

The kurtosis of the measured masses is close to 0 while the Fisher's coefficient of kurtosis of the radii is 1.1 meaning the tails are heavier than a normal distribution. This kurtosis is not further investigated in this work but is captured by using the samples of mass and radius from the posterior distributions.

In summary, we find among other results that when considering density uncertainties below approximately 10\% the degeneracy remains between core, mantle, and water in a three layer model and the overall likely wmf is unchanged by better observations. However, a two fold decrease in density uncertainty leads to a two fold decrease in the uncertainty of wmf at a given r-cmf. 

The science goals of the PLAnetary Transits and Oscillations of stars (PLATO) telescope is to reach radius uncertainties of 3\% for a planet around a star with V-magnitude of 10. Extreme precision radial velocity is expected to reach a 10\% mass uncertainty for Earths and Super-Earths on the inner edge of the habitable zone around G-type stars. Taking these values and assuming they are uncorrelated leads to a 13.5\% density---slightly larger than considered in this section. However, with as few as 3 transits around the brightest stars and more transits observed in the long duration fields over the planned 4-year mission, the planet radius uncertainty can reach as low as the star's radius uncertainty (often taken to be near 1\%) \citep{Rauer2024}. Mass uncertainties can reach lower than the target of 10\% around small stars and transit timing variations can further refine mass measurements in multiplanet systems. With small planet detections expected in the thousands with PLATO there will likely be a sizable sample of planets with density uncertainties under 5\% after follow-up observation.

\begin{table*}
	\centering
	\caption{2,500 samples of radius were generated for each of 4 skewed distributions and one non-skewed distribution and paired with with masses drawn from a Gaussian with 1.039$\pm$0.31 M$_\oplus$. Uncertainty reported as a percentage fo the median value in the column to the left. The Fisher-Pearson coefficient of skewness is reported in the ``skew'' column for the quantity in the column two to the left.}
	\addtolength{\tabcolsep}{-1pt}
	\begin{tabular}{ccccccccccccccc} 
		\hline
		Radius &   +1$\sigma$ \% & skew & Density &  +1$\sigma$ \% & skew  & Earth &  +1$\sigma$ \% & skew  & Mercury &  +1$\sigma$ \%  & skew  & cmf &  +1$\sigma$ \% & skew  \\
		R$_\oplus$ &   $-1\sigma$ \% & & $\rho_\oplus$ & $-1\sigma$ \% &  & wmf \% &   $-1\sigma$ \% & & wmf \% &   $-1\sigma$ \% & & \% &   $-1\sigma$ \% & \\
		
		\hline
	    1.045 & 1.1 & $-$1.2 & 0.91 & 5.9 & 1.0 & 7.0  & 41  & 0.1  & 21 & 13 & $-$0.6 & 14 & 57 & 0.8 \\
	          & 1.7 &  &          & 4.8 &      &      & 44  &      &    & 15 &     &    & 43 &    \\
	    1.045 & 1.3 & 0.0 & 0.91 & 5.0 & 0.2  & 7.3  & 40  & 0.3  & 21 & 13 & 0.1 & 13 & 54 & 0.2 \\
	          & 1.3 &     &      & 4.6 &      &      & 38  &      &    & 13 &     &    & 46 &   \\
	    1.045 & 1.7 & 0.9 & 0.91 & 4.9 & $-$0.3 & 7.1  & 52  & 1.0  & 21 & 17 & 0.8 & 14 & 43 & 0.0  \\
	          & 1.2 &  &      & 5.8 &        &      & 38  &      &    & 13 &     &    & 50 &       \\
	    1.044 & 2.3 & 1.7 & 0.91 & 4.6 & $-$1.0 & 6.9  & 67  & 1.8  & 21 & 21 & 1.7 & 15 & 33 & $-$0.1    \\
	          & 1.2 &  &      & 6.8 &       &          & 36  &      &    & 12 &     &    & 47 &    \\
	    1.045 & 3.4 & 2.5 & 0.90 & 5.3 & $-$1.5 & 7.4  & 91  & 3.0  & 21 & 30 & 2.8 & 15 & 33 & $-$0.0   \\
	          & 1.2 &  &      & 9.4 &       &          & 38  &      &    & 13 &     &    & 47 &   \\	          
		\hline
	\end{tabular}\label{tab:skew}
\end{table*}

\section{Model Uncertainties}
\label{sec:modelunc}

In Section \ref{sec:model}, we described the base model used to infer the wmf of T1-f. While our default model was chosen to include up-to-date EoSs, the range of possible interiors for terrestrial planets is uncertain. In this section, we test choices in our model parameters. Although the distinction is not strict, we separate model uncertainties in the following four sections from uncertainties in internal EoS parameters which are examined in the Section \ref{sec:eos}. We examine our assumptions of no atmosphere (\ref{sec:atm}), pure silicate mantles (\ref{sec:mant}), pure iron cores (\ref{sec:core}). We also discuss compilations of hydrosphere EoSs (\ref{sec:hydro}). Lastly, throughout we discuss effects of temperature and layers in thermal equilibrium.

\subsection{Atmosphere}\label{sec:atm}

In this work, we focus on three-layer interior solutions with a hydrosphere to T1-f's observables. However, there are solutions with a rocky mantle, iron core, and high mean molecular weight atmosphere. While water-vapor atmospheres would be condensed at T1-f's temperature \citep{Acuna2021}, we include here a brief analysis of ideal gas atmospheres with a mean molecular weight similar to that of Venus (44 g/mol) and Titan (28 g/mol). The upper portion of the atmosphere between a pressure of 100 mbar and 100 bar is isothermal at 218 K and the temperature increases adiabatically at pressures above 100 bar.  

\begin{figure}[ht!]
\plotone{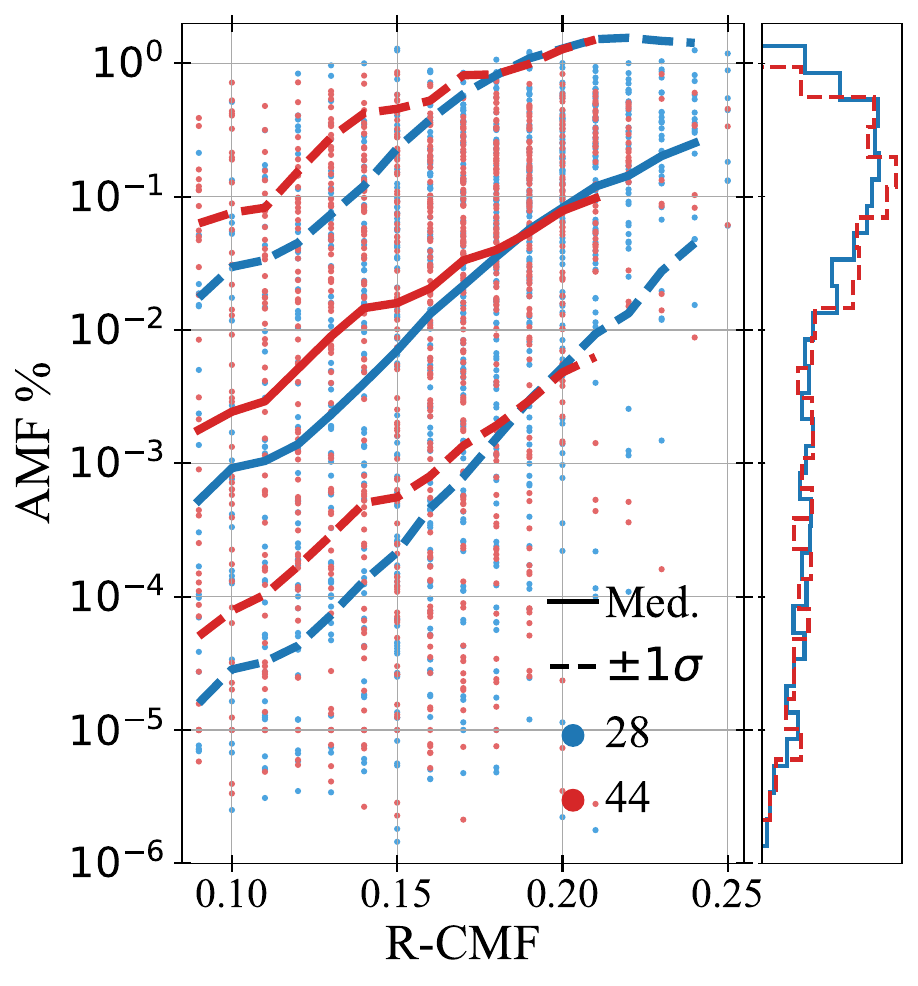}
\caption{The inferred amf of T1-f at 0.01 steps in r-cmf from 0.09 to 0.25 for atmospheres with mean molecular weight of 28 g/mol (\textit{blue}) and 44 g/mol (\textit{red}). Median and 1$\sigma$ bounds shown along with 10\% of the underlying data. \textit{Right}, a normalized histogram of amf. 
}
\label{fig:atm}
\end{figure}

We set a minimum r-cmf for these solutions of 0.09 below which a majority of draws of mass and radius need no atmosphere or hydrosphere. In addition, for a given mass and r-cmf of a planet, there is a maximum radius that can be reached by increasing the atmosphere mass fraction (amf) with our model while keeping total mass the same. For example, a 1.039 M$_\oplus$ planet reaches a maximum radius of 1.018 R$_\oplus$ at 2.8\% amf for an Earth-like r-cmf with a Venus-like mean molecular weight. Thus, modeling T1-f with a high mean-molecular weight ideal gas atmosphere requires a smaller r-cmf than Earth and nearly all solutions need a r-cmf lower than 0.25. 

Shown in Fig.~\ref{fig:atm}, we see that for a 28 g/mol molecular weight, the finder returns a median r-cmf of 0.17$^{+0.4}_{-0.05}$ and the amf is highly skewed with a median of 0.026\%. For a 44 g/mol molecular weight, the finder returns an r-cmf of 0.16$^{+0.3}_{-0.05}$ and the amf is highly skewed with a median of 0.022\%. 

If T1-f has negligible water, we find the most likely model of T1-f is a core half the mass of Earth's core (17\%) and an atmospheres over twice the mass of Venus's atmosphere (0.02\%). At the bottom of the atmosphere of this median solution the surface is at 306 K and 200 bar. However, atmosphere masses much smaller and much larger by an order of magnitude are within the 1$\sigma$ range. Larger cores and smaller atmospheres are possible by adding a 4th hydrosphere layer to the model adding further degeneracy to solutions. 

Atmospheres are best modeled by dedicated numerical methods that are coupled to interior models (e.g. \citealt{Acuna2023, Rigby2024}). Rather than our ideal gas atmosphere and temperature profile inspired by \citet{Nixon2021}, coupled models capture the transit pressure, temperature profile, and non-ideal behavior in the atmosphere. These effects provide a more realistic temperature and pressure for the top of the interior model. In the next two sections, we remove the atmosphere but provide insight into how changing the surface of the hydrosphere or mantle affects the interior.

\subsection{Mantle Assumptions}\label{sec:mant}

\begin{figure*}[ht!]
\plottwo{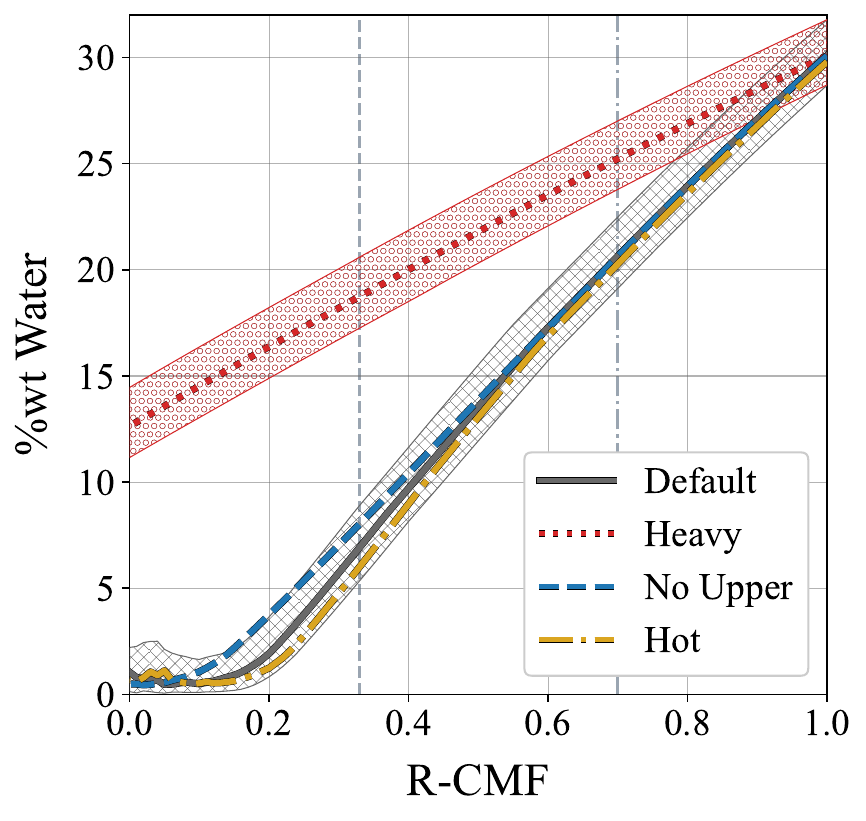}{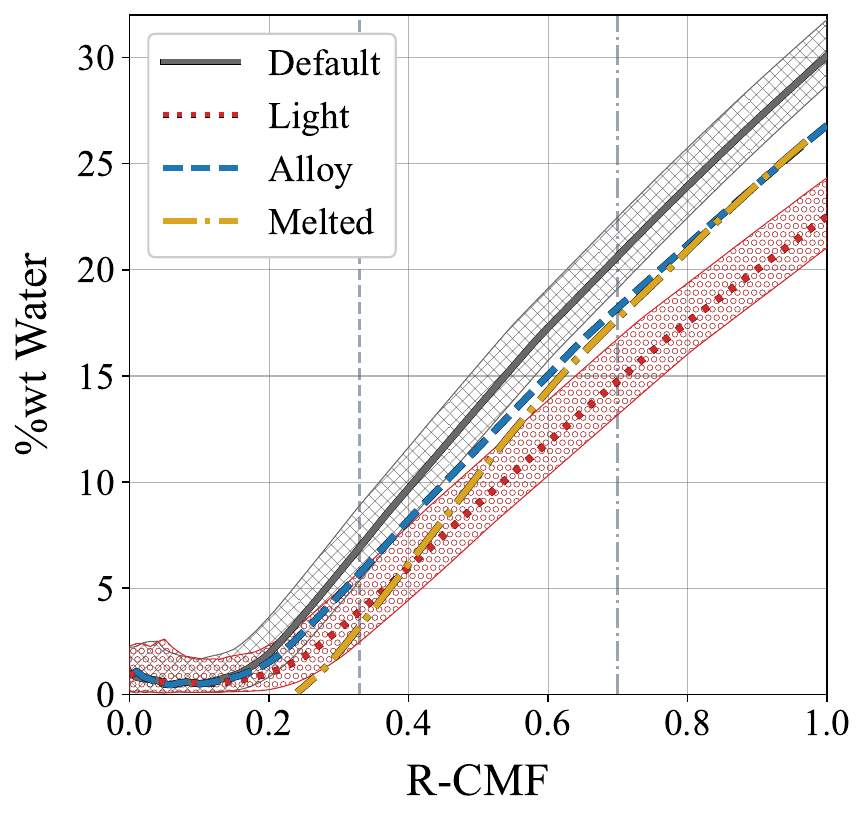}
\plottwo{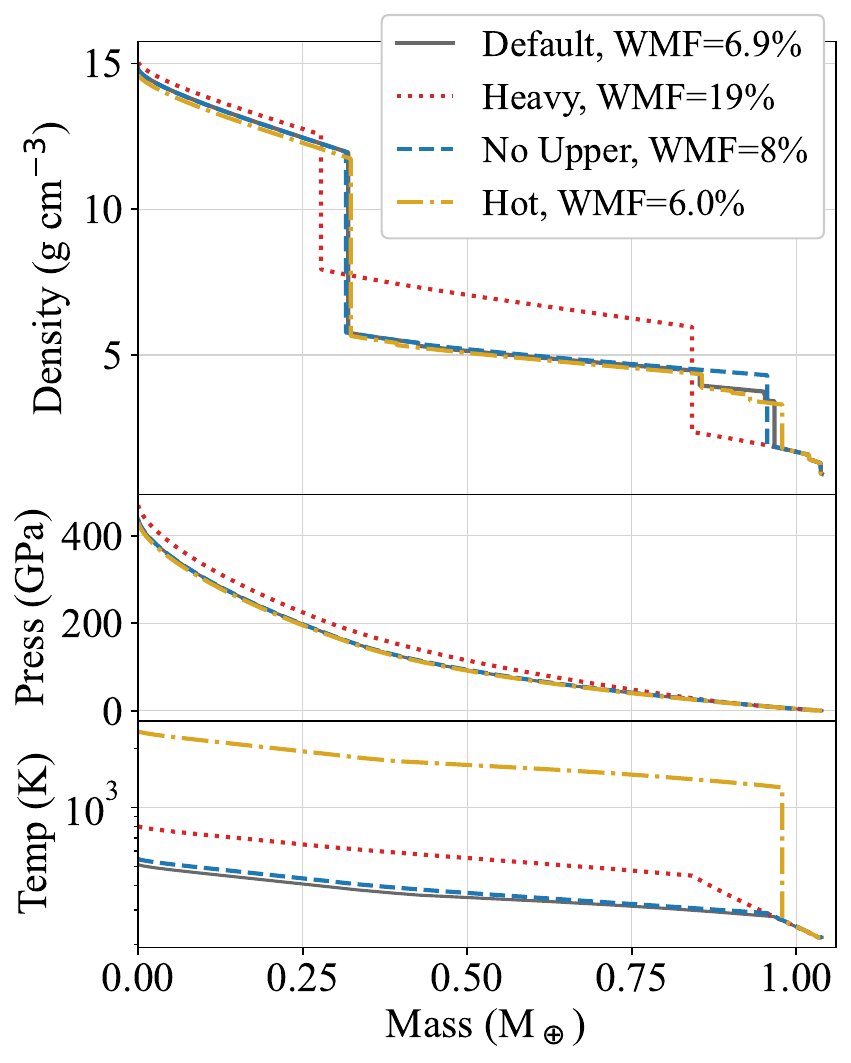}{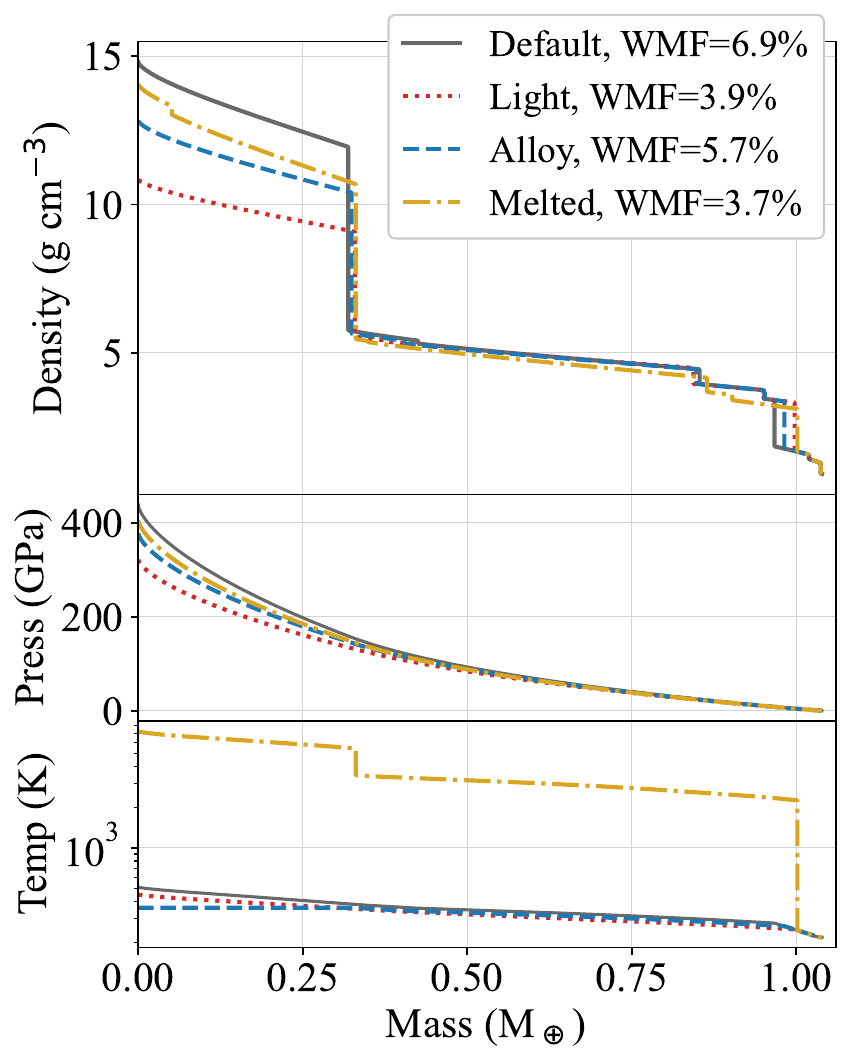}
\caption{\textit{Top}, wmf across r-cmf for four models with different mantle (\textit{left}) and core (\textit{right}) assumptions. \textit{Left}, 1-$\sigma$ bounds shown for default and ``heavy'' FeSiO$_3$ mantle and the median value is shown for models without an upper mantle and a ``hot'' mantle started at 1000 K. \textit{Right}, 1-$\sigma$ bounds shown for default and ``light'' FeS mantle and the median value is shown for models with a partially ``melted'' core and a FeSi ``alloy'' core. \textit{Bottom}, the density, pressure and temperature with enclosed mass for the interior of T1-f with a 33\% r-cmf and the median wmf with the seven models from the top. The FeSi alloy core EoS has no thermal parameters and thus temperature is plotted as constant in the core for this model. The compositional assumptions can cause large variations in inferred wmf while temperature assumptions in most cases have a smaller effect.
}
\label{fig:mantlecores}
\end{figure*}

We examine here our assumptions of mantle composition---simplified upper mantle, lack of iron silicates, and the temperature at the top of the mantle---on the inferred wmf. Our base model features an upper-mantle of olivine polymorphs which decompose to bridgmanite, an enstatite polymorph, at high-pressure. Other interior models (e.g. \citet{Unterborn2023}) find mantle equilibrium phases with the thermodynamics package \textsc{Perple\_X} \citep{Connolly2009}. However, this level of complexity often has little affect on the inferred bulk properties. If we ignore the upper mantle completely and model T1-f with only bridgmanite and post-perovskite magnesium silicate, the wmf is 8.0\% at Earth-like r-cmf shown in Fig.~\ref{fig:mantlecores} \textit{left}. This is a 1.07\% higher wmf than our base model from removing the less dense upper-mantle magnesium silicates. While ringwoodite is on average 18\%, wadsleyite is 20\%, and olivine is 25\% less dense than bridgmanite at pressures below 20 GPa, the upper mantle of T1-f only accounts for small portion of the planet under the hydrosphere. The pressure at the bottom of the hydrosphere of our Earth-like r-cmf solution is around 8.0 GPa. At r-cmf's above 0.5 the pressure at the bottom of the 14\% wmf hydrosphere is higher than the stability of upper mantle phases.

While the upper mantle's exact chemistry is a small effect, a larger assumption is our use of pure magnesium silicates in the lower mantle. While we could consider the effects of aluminum and calcium, the largest effect is expected to be iron because of its large molar mass. \citet{Yang2024} finds that iron in the mantle also form a perovskite structure at pressures above 63 GPa and that the phase had a similar cell volume and compressibility to that of bridgmanite (silicate perovskite). We model a iron perovskite mantle by keeping the same magnesium silicate EoSs, but increasing the molar mass to that of a iron silicate with no magnesium. The material is approximately 30\% denser than silicate perovskite at a given pressure and temperature. With a dense mantle, T1-f on average across all r-cmf needs 5.8\% more of its mass in a hydrosphere. The wmf across all r-cmf is shown in the left panel of Fig.~\ref{fig:mantlecores} along with the interior conditions of the median T1-f solution with a Earth-like r-cmf. If all of the iron is in the mantle with no core, the median T1-f has a 12.6\% wmf.

Another small affect for T1-f is the temperature at the top of the mantle. While the Earth has an effective temperature around 300 K the temperature of the mantle under the crust is at least 1000 K higher. \citet{Seager2007} showed that temperature has a small effect on a material's density at high pressure and if treating the planet as isothermal the planet's radius is underestimated by approximately 1.2\%. Subsequently a number of works ignore the effect of temperature (e.g. \citealt{Weeks2024}). For T1-f we find that starting the top of the mantle hot with a 1000 K discontinuity from the hydrosphere to mantle decreases the wmf at an Earth-like r-cmf by only 0.9\%. A 1500 K discontinuity decrease the wmf by 1.6\%. As shown in the left bottom panel of Fig.~\ref{fig:mantlecores}, this is a combined effect of a decrease in the density of the high-pressure phases and the stability of olivine up to moderately higher pressures in the upper mantle.

Without crossing a major phase boundary as explored with melts in the next section, the mantle temperature influences inferred wmf on the order of 1\%. While upper mantle chemistry has a similar influence, the largest model dependency is on the density of the mantle which in the extreme case can increase the inferred wmf by over 10\% at low r-cmfs. We stay near Earth-like mantle mineralogy and do not explore exotic mantles enriched in carbon bearing minerals. The mantle can also be made less dense with the inclusion of water \citep{Dorn2021,Vazan2022}, but the small mass of T1-f should limit the dissolution of water in rock to small amounts that are not explored here.

\subsection{Core Assumptions}\label{sec:core}

Our default core is pure iron in a hexagonal closed pack structure \citep{Smith2018}. We examine how changing the core composition affects the inferred wmf in two models. First the ``light'' model approximates a FeS core by changing the mean molecular weight of the iron to be that of 50\% sulfur and 50\% iron while keeping the iron's EoS properties. The 50\% sulfur core is thus equivalent to having a 43\% silicon, a 31\% oxygen, or a 22\% hydrogen by weight core. In addition to this simplified ``light'' model, we run an ``alloy'' model with an EoS from \citet{Wicks2018} for an iron-silicon alloy with 15\% weight silicon. They find that this wt\% of silicon transforms the alloy from hexagonal close-packed to a body-centered cubic structure which changes its response to pressure.

Across all r-cmf shown in Fig.~\ref{fig:mantlecores} \textit{right}, the light core model averages a wmf 4.7\% less than our default model and 3\% less at an Earth-like r-cmf. The change in wmf between the two models increases with increasing r-cmf with a Mercury-like r-cmf having 5.9\% less water than the default. The alloy core requires a 2\% smaller wmf across all r-cmf and 1.2\% less at Earth-like r-cmf growing to 2.4\% less at a Mercury-like r-cmf.

The temperature profile within the planet's core remains uncertain. For the Earth, the fact that the outer core is liquid sets a lower limit on the temperature at its outer boundary. This high temperature suggests that, at Earth's core–mantle boundary, the temperature rises sharply---by approximately 1000 K---within a thin region known as the thermal boundary layer \citep{Lay2008, Valencia2007, Nomura2014}. Additionally, whether the temperature follows an isentropic profile in the core remains a subject of debate, due in part to uncertainties in the thermal conductivity of iron \citep{yin2022electrical}. High thermal conductivity values obtained from \textit{ab initio} calculations suggest the presence of a thermally stratified layer at the top of the liquid core, potentially extending up to a thousand kilometers in thickness \citep{2012Natur.485..355P}. Within this stratified layer, the temperature gradient may be reduced to roughly one-third of the adiabatic gradient, implying a lower overall core temperature.

To explore the range of possible core structures for T1-f, we modified the default adiabatic temperature profile by introducing a 2000 K temperature discontinuity between the hydrosphere and mantle---leading to silicate melting---and another 2000 K discontinuity between the mantle and core, representing the upper bound of plausible temperatures. This creates a layer of liquid iron in all of the planets. This extreme temperature model needs 3.2\% smaller wmf at an Earth-like r-cmf and 2.9\% smaller wmf at a Mercury-like r-cmf. With an Earth-like r-cmf 84.6\% of the core's mass is melted and with a Mercury-like r-cmf 78.5\% of the core's mass remains melted.

For an Earth-like core-to-mantle ratio, the model assumptions of the core's composition influence the inferred wmf up to 3\%. For a small planet, a liquid core has a similar effect to a large amount of light elements in the core. At larger cmfs, the weight of lighter elements in the core becomes the dominant assumption to inferred wmf in our model.

\subsection{Hydrosphere Assumptions} \label{sec:hydro}

Our default model for the hydrosphere includes a compilation of several EoS for the many phases of H$_2$O and assumes that the surface temperature is below the freezing point. However, T1-f is within the habitable zone \citep{TRAPPIST1} where liquid water can exist on the surface with only slight heating from a thin atmosphere.

\begin{figure*}[ht!]
\plottwo{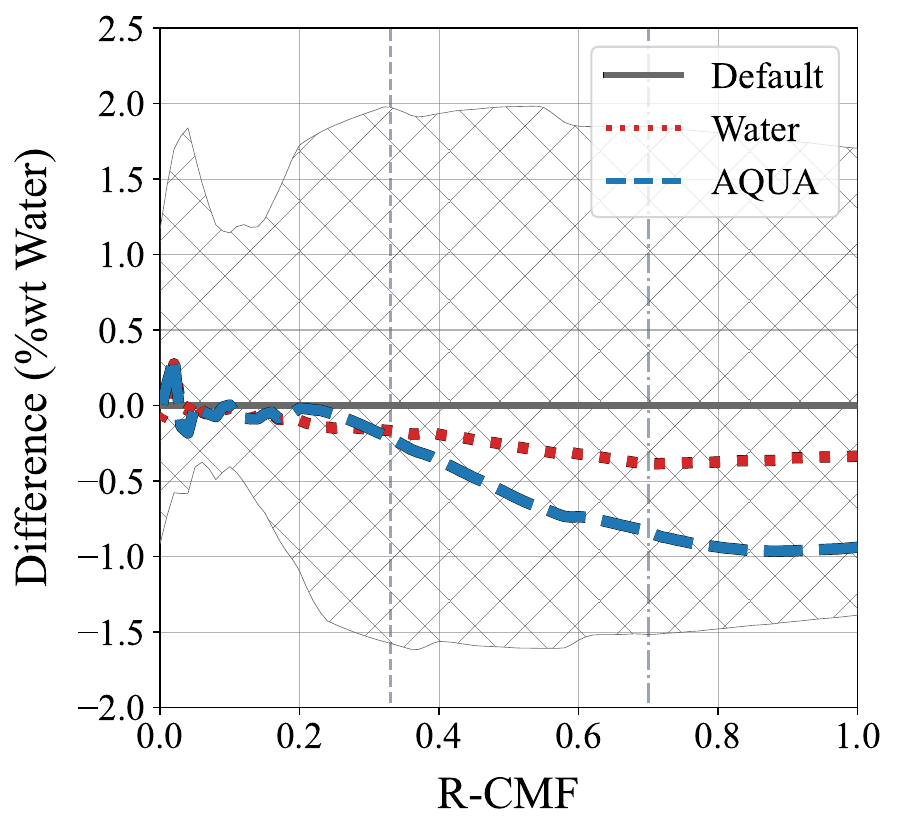}{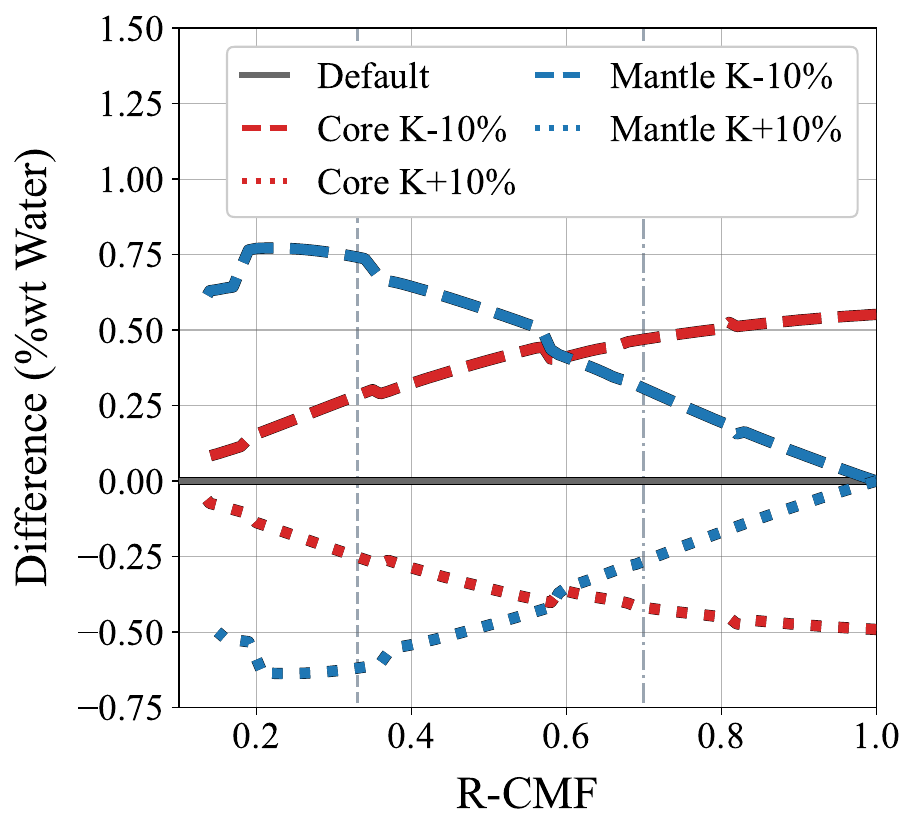}
\caption{\textit{Left}, difference in wmf between two different hydrosphere assumptions and the default model. \textit{Red dotted} is models with the surface temperature raised to create a layer of liquid water instead of ice Ih. \textit{Blue dashed} is models with the AQUA compilation of EoSs from \citet{Haldemann2020}. The \textit{hatched} region shows the 1-$\sigma$ region of wmf from observational uncertainties with our default model. \textit{Right}, difference in wmf between planet models with different bulk moduli for the mantle or core and our default planet model. In \textit{red}, the bulk modulus of the core is increased (\textit{dotted}) and decreased (\textit{dashed}) by 10\%. In \textit{blue}, the bulk modulus of the bridgmanite and post perovskite is increased/decreased by 10\%. 
}
\label{fig:hydrobulk}
\end{figure*}

Fig.~\ref{fig:hydrobulk} \textit{left}, shows the difference in inferred wmf with r-cmf if the surface temperature is raised to 274 K above the melting point of H$_2$O. Although, as expected, the surface has a higher density when assuming water over ice Ih, the increase in temperature in the interior decreases the overall density, leading to a reduction in the inferred wmf. Since the temperature gradient is greatest in the water shell, the difference in inferred wmf increases with r-cmf and higher inferred wmfs. Yet, at all r-cmf, the difference remains smaller than 0.5\%.

We also compare the default compilation of EoSs in \magra\ for the hydrosphere to that of the AQUA collection from \citet{Haldemann2020}. The collection uses all different measurements for phases of H$_2$O than the default compilation. To work with AQUA, we implement into \magra\ a method to use $rho$-P-T-$dT/dP$ tables for EoS using a linear interpolator. We show the difference in the inferred wmf between the two compilations in Fig.~\ref{fig:hydrobulk} \textit{left}. The inferred wmf agrees to within 1\% between the EoS compilations. The uncertainty between the water and high-pressure ice EoS from \magra's EoSs and AQUA is comparable to changes in mantle temperature from Section \ref{sec:mant}.

Although the water phase diagram and the behavior of each phase under pressure and temperature are still not fully understood, for a small planets with a wmf below 10\% in a condensed phase, differences in experimental data have little impact on the overall bulk properties. However, details of the adiabatic temperature gradient in the water-ice layer can still create significantly different interior and surface conditions for future habitability studies.

\section{Bulk Modulus Uncertainties in the Equation of State}
\label{sec:eos}

The median interior model for T1-f includes 11 different materials with on average 10 parameters defining how each material responds to pressure and temperature. While the previous section deals with assumptions in our model and with hydrosphere EoS compilations, we address here how a change in one of these parameters can affect our inferences. We focus on the bulk modulus (K) which is the resistance of a material to compression and the largest factor in determining how a material's density changes within the interior.

In \magra, we store a library of EoSs from the literature. We have bulk modulus for high pressure iron from different experimental and theory works ranging from 148 GPa from \citet{Dorogokupets17} using a Vinet fitting formula to 254 GPa from \citet{Bouchet13} using a Holzapfel fit. \citet{Dorogokupets17} finds that when comparing measurements using the same fit the bulk modulus falls in a range of $\pm$2 GPa, although they note a couple of anomalous measurements. The recent measurement by \citet{Smith2018} using a Vinet fit falls significantly outside of this report $\pm$2 GPa uncertainty with a higher bulk modulus of 177.7 GPa. 

Mantle EoS measurements have similar discrepancies based on limited data. For bridgmanite our library includes a 3rd order Birch-Murnaghan EoS from \citet{Shim2000} with a bulk modulus of 261 GPa and a Vinet-fit from \citet{Oganov04} with bulk modulus of 230 GPa. A Vinet-fit for the same material will give a higher density than a BM3 fit at high pressures, and the smaller bulk modulus found in \citet{Oganov04} only further increases the density at a given pressure. For post-perovskite, \citet{Sakai16} reports a bulk modulus of 203 GPa with a Keane EoS and \citet{Dorogokupets2015} reports a higher 253.7 GPa with a 3rd order Birch-Murnaghan EoS. 

With the measurements discussed above in mind, we simplify our analysis to considering a $\pm$10\% change in the bulk modulus of either the core or mantle materials. A 10\% change in this important quantity seems to be plausible as more data and better fitting approximations are achieved. In the core we consider the Vinet fit from \citet{Smith2018} and vary the bulk modulus from 159.93 GPa to 195.47 GPa. In the mantle, we change the bulk modulus of both the bridgmanite and post perovskite in the same direction. The bridgmanite varies from 207 to 253 GPa and the post-perovskite varies from 182.7 to 223.3 GPa. 

The result to the inferred wmf is shown in Fig.~\ref{fig:hydrobulk} \textit{right}. A lower bulk modulus leads to a squishier interior and the need for a large wmf to offset the higher density interior. Visa-versa, a higher bulk modulus leads to a lower inferred wmf.  When the mantle is largest with hydrosphere, a 10\% change in the bulk modulus of the mantle materials leads to a +0.75\% -0.6\% difference in wmf. When there is no mantle, a 10\% change in the bulk modulus of the core material leads to a $\pm$0.5\% difference in the wmf.

\section{Conclusion}
\label{sec:conclusion}

In this work, we explored the effects of observational, model, and experimental uncertainties on inferring the water mass fraction (wmf) from mass and radius measurements for small planets. We use as a benchmark observation TRAPPIST-1 f (T1-f), a nearly Earth-mass and Earth-radius planet situated in the habitable zone. Using the open-source interior solver \magra, we examined T1-f's interior structure and composition.

Our results show that the outer five TRAPPIST-1 planets likely possess significant hydrospheres. Focusing on T1-f we find an inferred wmf of 16.2\% ± 9.9\%, depending the core to mantle mass ratio. At an Earth-like refractory core mass fraction, r-cmf of 0.33, the wmf is constrained to 6.9\% $\pm$ 2.0\%. These values are sensitive to uncertainties in the planet's bulk density, with higher precision in correlated mass-radius measurements significantly reducing inferred wmf uncertainty. Using correlated mass-radius measurements and capturing the skew of the data creates smaller and more realistic uncertainties than fitting Gaussian to reported values.

We explored the influence of model assumptions, including mantle composition, core composition, and hydrosphere temperature, on inferences of T1-f's composition. Some key results:

\begin{itemize}
    \item Variations in mantle mineralogy (e.g. adding iron-rich perovskite) increases the inferred wmf by a maximum of approximately 10\%, particularly at low r-cmfs. 
    \item Changes to the core composition, such as light element enrichment or partial melting, reduce the required wmf by up to 3–5\%, particularly for high r-cmf.
    \item Changing the temperature at the top of the mantle can increase or decrease the inferred wmf by 1–2\% as long as no melting occurs.
    \item Hydrosphere assumptions, such as liquid versus frozen surface water, affect wmf inferences by less than 0.5\%, indicating that phase transitions in the water layer are secondary to mantle and core density considerations.
    \item Additionally, experimental uncertainties in the bulk modulus of mantle and core materials introduce \textless0.75\% deviations in wmf.
\end{itemize}

Considering all sources of uncertainty together, the total uncertainty in wmf at a given r-cmf spans approximately ±10\% larger than the $\pm$2\% from observations. Model assumptions, particularly core and mantle properties, introduce uncertainties that often exceed those from observational uncertainty. Currently, model assumptions dominate over laboratory constraints, but in the future laboratory measurements become important when density uncertainties are below 1\%. Future high-precision observations, combined with advances in understanding the range of likely mantle and core compositions, will be crucial in reducing these uncertainties and unlocking the true nature of small planet interiors.

To improve constraints on exoplanet interiors, future modeling efforts should first use realistic samples of density, mass, and radius. Additionally, observational works should make available their mass and radius posterior samples and report the uncertainty in density or surface gravity (see also \citealt{RodriguezMartinez2021}). Second, incorporate more diverse mantle and core compositions and vary the temperature profiles in the interior. Works should then systematically evaluate these model inter-dependencies. 


Although planetary interiors cannot be directly observed, upcoming observations are expected to provide valuable insights into planetary composition and a planet's mass distribution within the interior. The improved accuracy of the mass-radius relation in large exoplanet databases, facilitated by PLATO \citep{Rauer2024}, will help reduce some of the uncertainties in demographic composition, as discussed in Section \ref{sec:obs}, and further tie planet compositions to host star compositions \citep{Putirka2019,Adibekyan2021,Brinkman2024}. Future spectroscopic observations by space missions such as JWST \citep{TRAPPIST-1JWSTCommunityInitiative2024} and ARIEL \citep{Tinetti2018} will indicate atmospheric compositions and their abundances, offering constraints on interior composition through the theory of equilibrium chemistry and outgassing \citep{Madhusudhan2021,Tsai2021,Shorttle2024,Brachmann2025}. Additionally, if and when the Love number, $k_2$, can be measured, it will provide crucial insights into the internal density distribution \citep{Kellermann2018,Padovan2018,Baumeister2023}, further constraining possible interior compositions and distribution.

Beyond observational advancements, improvements in material science through laboratory and numerical experiments will enhance planetary interior modeling. Better-constrained bulk modulus values for planetary building blocks will refine mass-radius models, as demonstrated in Section \ref{sec:eos}. Furthermore, a better understanding of element miscibility in mixtures will enhance interior structure modeling by clarifying the conditions under which differentiation can occur (e.g. \citealt{Vazan2022, Luo2024}).

TRAPPIST-1 f remains a prime target for understanding small exoplanet interiors. While improved observations will continue to refine mass-radius constraints, theoretical and experimental advancements in planetary materials are equally necessary to break degeneracies in planetary composition. This study provides a framework for systematically assessing these uncertainties, paving the way for more robust interpretations of small exoplanets.

\begin{acknowledgments}
The authors thank an anonymous referee for strengthening this work. We appreciate DRR's dissertation committee members for stimulating discussions. We thank Lorena Acu\~na and team for use of their data. We acknowledge support from the College of Sciences and the Star and Planet Formation Group at the University of Nevada, Las Vegas (UNLV).  All simulations were supported by the Cherry Creek Cluster at UNLV.  We acknowledge that the study resulting in this publication was assisted by the President's UNLV Foundation Graduate Research Fellowship. A.V. acknowledges support by ISF grants
770/21 and 773/21.
 C.H. is sponsored by Shanghai Pujiang Program (grant NO. 23PJ1414900). 
\end{acknowledgments}

%



\software{\textsc{python-ternary} \citep{ternaryplot}, 
    \textsc{NumPy} \citep{numpy}, 
    \textsc{Matplotlib} \citep{matplotlib}, 
    \textsc{cmasher} \citep{cmasher}
          }



\appendix

\section{Data Representations}\label{app:plots}

Data visualizations are important for representing the many solutions to a planet's interior for a single observation of mass and radius. Here, we go into further details on data visualizations that we use in this work and their relationship to previous works.

Fig.~\ref{fig:T1dh} show a ternary plot which represents the parameter space defined by three variables which sum to a constant. Each point on the equilateral triangle represents a unique combination of three components. Each vertex of the equilateral triangle represents 100\% of one component and 0\% of the other two components. On the opposite edge from a vertex the component that is 100\% at that vertex is not-present (0\%). The percent of the component increases perpendicular to that edge. At a point the percentage of a component is given by the ratio of the length of a line perpendicular from an edge to the point to the height of the triangle. To make ternary plots, we use the package python-ternary by \citet{ternaryplot}.

In this work, we use water-mantle-core mass fraction ternary diagrams where the axis are read counterclockwise from the apex. The position of a point gives a unique percentage of mass in each fully-differentiated layer of a planet. However, the axis of a ternary may be rotated which makes comparing ternaries more difficult.  In addition, the axis values have a valid clockwise and counterclockwise orientations for their labels.  

The water-mantle-core orientation plots unphysical, 100 per cent water planets on the top vertex. This makes the independent mass ratio of core to mantle vary across the bottom edge while the dependent water mass percentage varies perpendicular to the bottom edge. Rocky planets with no significant water, like the Earth with 33\% core, plot on the bottom edge. Planets with no core plot on the left edge.

The orientation described here is the same as in \citet{Zeng2008} and \citet{Daspute2025}, although they place labels along the edges. Labels along the edges are common in the exoplanetary literature, but are difficult to read because percentages do not increase along lines orthogonal to the axis as in a typical two dimensional plot. Labels at the corner indicate that a quantity increases perpendicularly from the opposite edge toward that corner. \citet{Rogers2010}, \citet{Batalha2011}, \citet{Suissa2018}, \citet{Acuna2021}, and \citet{Haldemann2022} also use labels along the edge and use a mantle-water-core orientation. \citet{Brugger2016} and \citet{Valencia2007tern} use corner labels but also use the mantle-water-core orientation. This orientation is flipped horizontally and rotated counter-clockwise 120 degrees from our orientation. Furthermore, \citet{Unterborn2019} plots exoplanet host stars on a Si-Fe-Mg ternary, \citet{Madhusudhan2012} features a carbon-iron-MgSiO$_3$ ternary, and \citet{Neil2020} plots planets on a ``evaporated core''-``intrinsically rocky''-``gaseous'' planet ternary.

Planet's which have the same ratio of two layers plot on a line going from a vertex to the opposite edge. Ganymede plots to the left of the line running from Earth on the mantle:core axes to the apex on Fig.~\ref{fig:T1dh}. This is because the moon of Saturn likely is less differentiated than the Earth or has less iron to match gravity field measurements from satellites \citep{Schubert2007}. In Eq.~\ref{eq:RCMF}, we define refractory core mass fraction, r-cmf, which is constant along a line from the bottom edge to the apex. The r-cmf is 0 on the left edge and 1 on the right edge. We show an Earth-like 0.33 r-cmf line and a Mercury-like 0.7 r-cmf line on Fig.~\ref{fig:T1dh}.

If we take the r-cmf as the independent variable, we can transform the ternary into a more typical two dimensional plot with r-cmf on the x-axis and water mass fraction (wmf) on the y-axis. This plot is shown in Fig.~\ref{fig:T1all} and subsequent figures. There is no forbidden regions on the plot unlike a wmf versus cmf plot or for volume fractions as done in \citet{Dobos2019}. The r-cmf for our model of magnesium silicate mantles and iron cores can be equated to $\frac{f_{\text{Fe}}}{xf_{\text{Si}}+f_{\text{Fe}}}$ where $f$ is the weight percentage of the elements and $x$ is the inverse of the silicon weight percentage of the mantle.

In \citealt{Magrathea}, Fig.~6 shows a water-mantle-core ternary with all of the radii a three-layer planet of a given mass can have at each position on the plot. An additional constraint of planet radius gives us a curve of three-layer mass fractions which have the same radius. For three layers, an increase in the core mass will lower the radius, but that radius can stay the same if more of the mantle's mass is also converted to hydrosphere. Thus a mass and radius constraint gives a curve on both the ternary and r-cmf plots. Our process of finding these curves and using them to constrain a planet's interior structure is described in Section \ref{sec:finder}.


\bibliography{interiors}{}

\begin{thebibliography}{}
\expandafter\ifx\csname natexlab\endcsname\relax\def\natexlab#1{#1}\fi
\providecommand{\url}[1]{\href{#1}{#1}}
\providecommand{\dodoi}[1]{doi:~\href{http://doi.org/#1}{\nolinkurl{#1}}}
\providecommand{\doeprint}[1]{\href{http://ascl.net/#1}{\nolinkurl{http://ascl.net/#1}}}
\providecommand{\doarXiv}[1]{\href{https://arxiv.org/abs/#1}{\nolinkurl{https://arxiv.org/abs/#1}}}

\bibitem[{{Acu{\~n}a} {et~al.}(2023){Acu{\~n}a}, {Deleuil}, \& {Mousis}}]{Acuna2023}
{Acu{\~n}a}, L., {Deleuil}, M., \& {Mousis}, O. 2023, \aap, 677, A14, \dodoi{10.1051/0004-6361/202245736}

\bibitem[{{Acu{\~n}a} {et~al.}(2021){Acu{\~n}a}, {Deleuil}, {Mousis}, {Marcq}, {Levesque}, \& {Aguichine}}]{Acuna2021}
{Acu{\~n}a}, L., {Deleuil}, M., {Mousis}, O., {et~al.} 2021, \aap, 647, A53, \dodoi{10.1051/0004-6361/202039885}

\bibitem[{{Adibekyan} {et~al.}(2021){Adibekyan}, {Dorn}, {Sousa}, {Santos}, {Bitsch}, {Israelian}, {Mordasini}, {Barros}, {Delgado Mena}, {Demangeon}, {Faria}, {Figueira}, {Hakobyan}, {Oshagh}, {Soares}, {Kunitomo}, {Takeda}, {Jofr{\'e}}, {Petrucci}, \& {Martioli}}]{Adibekyan2021}
{Adibekyan}, V., {Dorn}, C., {Sousa}, S.~G., {et~al.} 2021, Science, 374, 330, \dodoi{10.1126/science.abg8794}

\bibitem[{{Agol} {et~al.}(2021){Agol}, {Dorn}, {Grimm}, {Turbet}, {Ducrot}, {Delrez}, {Gillon}, {Demory}, {Burdanov}, {Barkaoui}, {Benkhaldoun}, {Bolmont}, {Burgasser}, {Carey}, {de Wit}, {Fabrycky}, {Foreman-Mackey}, {Haldemann}, {Hernandez}, {Ingalls}, {Jehin}, {Langford}, {Leconte}, {Lederer}, {Luger}, {Malhotra}, {Meadows}, {Morris}, {Pozuelos}, {Queloz}, {Raymond}, {Selsis}, {Sestovic}, {Triaud}, \& {Van Grootel}}]{Agol2021}
{Agol}, E., {Dorn}, C., {Grimm}, S.~L., {et~al.} 2021, PSJ, 2, 1, \dodoi{10.3847/PSJ/abd022}

\bibitem[{{Aguichine} {et~al.}(2021){Aguichine}, {Mousis}, {Deleuil}, \& {Marcq}}]{Aguichine2021}
{Aguichine}, A., {Mousis}, O., {Deleuil}, M., \& {Marcq}, E. 2021, \apj, 914, 84, \dodoi{10.3847/1538-4357/abfa99}

\bibitem[{{Barros} {et~al.}(2022){Barros}, {Demangeon}, {Alibert}, {Leleu}, {Adibekyan}, {Lovis}, {Bossini}, {Sousa}, {Hara}, {Bouchy}, {Lavie}, {Rodrigues}, {Gomes da Silva}, {Lillo-Box}, {Pepe}, {Tabernero}, {Zapatero Osorio}, {Sozzetti}, {Su{\'a}rez Mascare{\~n}o}, {Micela}, {Allende Prieto}, {Cristiani}, {Damasso}, {Di Marcantonio}, {Ehrenreich}, {Faria}, {Figueira}, {Gonz{\'a}lez Hern{\'a}ndez}, {Jenkins}, {Lo Curto}, {Martins}, {Micela}, {Nunes}, {Pall{\'e}}, {Santos}, {Rebolo}, {Seager}, {Twicken}, {Udry}, {Vanderspek}, \& {Winn}}]{Barros2022}
{Barros}, S.~C.~C., {Demangeon}, O.~D.~S., {Alibert}, Y., {et~al.} 2022, \aap, 665, A154, \dodoi{10.1051/0004-6361/202244293}

\bibitem[{{Batalha} {et~al.}(2011){Batalha}, {Borucki}, {Bryson}, {Buchhave}, {Caldwell}, {Christensen-Dalsgaard}, {Ciardi}, {Dunham}, {Fressin}, {Gautier}, {Gilliland}, {Haas}, {Howell}, {Jenkins}, {Kjeldsen}, {Koch}, {Latham}, {Lissauer}, {Marcy}, {Rowe}, {Sasselov}, {Seager}, {Steffen}, {Torres}, {Basri}, {Brown}, {Charbonneau}, {Christiansen}, {Clarke}, {Cochran}, {Dupree}, {Fabrycky}, {Fischer}, {Ford}, {Fortney}, {Girouard}, {Holman}, {Johnson}, {Isaacson}, {Klaus}, {Machalek}, {Moorehead}, {Morehead}, {Ragozzine}, {Tenenbaum}, {Twicken}, {Quinn}, {VanCleve}, {Walkowicz}, {Welsh}, {Devore}, \& {Gould}}]{Batalha2011}
{Batalha}, N.~M., {Borucki}, W.~J., {Bryson}, S.~T., {et~al.} 2011, \apj, 729, 27, \dodoi{10.1088/0004-637X/729/1/27}

\bibitem[{{Baumeister} {et~al.}(2020){Baumeister}, {Padovan}, {Tosi}, {Montavon}, {Nettelmann}, {MacKenzie}, \& {Godolt}}]{Baumeister2020}
{Baumeister}, P., {Padovan}, S., {Tosi}, N., {et~al.} 2020, \apj, 889, 42, \dodoi{10.3847/1538-4357/ab5d32}

\bibitem[{{Baumeister} \& {Tosi}(2023)}]{Baumeister2023}
{Baumeister}, P., \& {Tosi}, N. 2023, \aap, 676, A106, \dodoi{10.1051/0004-6361/202346216}

\bibitem[{{Beard} {et~al.}(2022){Beard}, {Robertson}, {Kanodia}, {Lubin}, {Ca{\~n}as}, {Gupta}, {Holcomb}, {Jones}, {Libby-Roberts}, {Lin}, {Mahadevan}, {Stef{\'a}nsson}, {Bender}, {Blake}, {Cochran}, {Endl}, {Everett}, {Ford}, {Fredrick}, {Halverson}, {Hebb}, {Li}, {Logsdon}, {Luhn}, {McElwain}, {Metcalf}, {Ninan}, {Rajagopal}, {Roy}, {Schutte}, {Schwab}, {Terrien}, {Wisniewski}, \& {Wright}}]{Beard2022}
{Beard}, C., {Robertson}, P., {Kanodia}, S., {et~al.} 2022, \apj, 936, 55, \dodoi{10.3847/1538-4357/ac8480}

\bibitem[{{Bezacier} {et~al.}(2014){Bezacier}, {Journaux}, {Perrillat}, {Cardon}, {Hanfland}, \& {Daniel}}]{Bezacier2014}
{Bezacier}, L., {Journaux}, B., {Perrillat}, J.-P., {et~al.} 2014, The Journal of Chemical Physics, 141, 104505, \dodoi{10.1063/1.4894421}

\bibitem[{{Bolmont} {et~al.}(2017){Bolmont}, {Selsis}, {Owen}, {Ribas}, {Raymond}, {Leconte}, \& {Gillon}}]{Bolmont2017}
{Bolmont}, E., {Selsis}, F., {Owen}, J.~E., {et~al.} 2017, \mnras, 464, 3728, \dodoi{10.1093/mnras/stw2578}

\bibitem[{{Bouchet} {et~al.}(2013){Bouchet}, {Mazevet}, {Morard}, {Guyot}, \& {Musella}}]{Bouchet13}
{Bouchet}, J., {Mazevet}, S., {Morard}, G., {Guyot}, F., \& {Musella}, R. 2013, Physical Review B, 87, 094102, \dodoi{10.1103/PhysRevB.87.094102}

\bibitem[{{Brachmann} {et~al.}(2025){Brachmann}, {Noack}, {Baumeister}, \& {Sohl}}]{Brachmann2025}
{Brachmann}, C., {Noack}, L., {Baumeister}, P.~A., \& {Sohl}, F. 2025, \icarus, 429, 116450, \dodoi{10.1016/j.icarus.2024.116450}

\bibitem[{{Brinkman} {et~al.}(2024){Brinkman}, {Polanski}, {Huber}, {Weiss}, {Valencia}, \& {Plotnykov}}]{Brinkman2024}
{Brinkman}, C.~L., {Polanski}, A.~S., {Huber}, D., {et~al.} 2024, \aj, 168, 281, \dodoi{10.3847/1538-3881/ad82eb}

\bibitem[{{Brugger} {et~al.}(2017){Brugger}, {Mousis}, {Deleuil}, \& {Deschamps}}]{Brugger2017}
{Brugger}, B., {Mousis}, O., {Deleuil}, M., \& {Deschamps}, F. 2017, \apj, 850, 93, \dodoi{10.3847/1538-4357/aa965a}

\bibitem[{{Brugger} {et~al.}(2016){Brugger}, {Mousis}, {Deleuil}, \& {Lunine}}]{Brugger2016}
{Brugger}, B., {Mousis}, O., {Deleuil}, M., \& {Lunine}, J.~I. 2016, \apjl, 831, L16, \dodoi{10.3847/2041-8205/831/2/L16}

\bibitem[{{Cadieux} {et~al.}(2022){Cadieux}, {Doyon}, {Plotnykov}, {H{\'e}brard}, {Jahandar}, {Artigau}, {Valencia}, {Cook}, {Martioli}, {Vandal}, {Donati}, {Cloutier}, {Narita}, {Fukui}, {Hirano}, {Bouchy}, {Cowan}, {Gonzales}, {Ciardi}, {Stassun}, {Arnold}, {Benneke}, {Boisse}, {Bonfils}, {Carmona}, {Cort{\'e}s-Zuleta}, {Delfosse}, {Forveille}, {Fouqu{\'e}}, {Gomes da Silva}, {Jenkins}, {Kiefer}, {K{\'o}sp{\'a}l}, {Lafreni{\`e}re}, {Martins}, {Moutou}, {do Nascimento}, {Ould-Elhkim}, {Pelletier}, {Twicken}, {Bouma}, {Cartwright}, {Darveau-Bernier}, {Grankin}, {Ikoma}, {Kagetani}, {Kawauchi}, {Kodama}, {Kotani}, {Latham}, {Menou}, {Ricker}, {Seager}, {Tamura}, {Vanderspek}, \& {Watanabe}}]{Cadieux2022}
{Cadieux}, C., {Doyon}, R., {Plotnykov}, M., {et~al.} 2022, \aj, 164, 96, \dodoi{10.3847/1538-3881/ac7cea}

\bibitem[{Chandra {et~al.}(2001)Chandra, Dagum, Kohr, Menon, Maydan, \& McDonald}]{openmp}
Chandra, R., Dagum, L., Kohr, D., {et~al.} 2001, Parallel programming in OpenMP (Morgan kaufmann)

\bibitem[{{Chaturvedi} {et~al.}(2022){Chaturvedi}, {Bluhm}, {Nagel}, {Hatzes}, {Morello}, {Brady}, {Korth}, {Molaverdikhani}, {Kossakowski}, {Caballero}, {Guenther}, {Pall{\'e}}, {Espinoza}, {Seifahrt}, {Lodieu}, {Cifuentes}, {Furlan}, {Amado}, {Barclay}, {Bean}, {B{\'e}jar}, {Bergond}, {Boyle}, {Ciardi}, {Collins}, {Collins}, {Esparza-Borges}, {Fukui}, {Gnilka}, {Goeke}, {Guerra}, {Henning}, {Herrero}, {Howell}, {Jeffers}, {Jenkins}, {Jensen}, {Kasper}, {Kodama}, {Latham}, {L{\'o}pez-Gonz{\'a}lez}, {Luque}, {Montes}, {Morales}, {Mori}, {Murgas}, {Narita}, {Nowak}, {Parviainen}, {Passegger}, {Quirrenbach}, {Reffert}, {Reiners}, {Ribas}, {Ricker}, {Rodriguez}, {Rodr{\'\i}guez-L{\'o}pez}, {Schlecker}, {Schwarz}, {Schweitzer}, {Seager}, {Stef{\'a}nsson}, {Stockdale}, {Tal-Or}, {Twicken}, {Vanaverbeke}, {Wang}, {Watanabe}, {Winn}, \& {Zechmeister}}]{Chaturvedi2022}
{Chaturvedi}, P., {Bluhm}, P., {Nagel}, E., {et~al.} 2022, \aap, 666, A155, \dodoi{10.1051/0004-6361/202244056}

\bibitem[{{Childs} {et~al.}(2023){Childs}, {Shakespeare}, {Rice}, {Yang}, \& {Steffen}}]{Childs2023}
{Childs}, A.~C., {Shakespeare}, C., {Rice}, D.~R., {Yang}, C.-C., \& {Steffen}, J.~H. 2023, \mnras, 524, 3749, \dodoi{10.1093/mnras/stad2110}

\bibitem[{{Connolly}(2009)}]{Connolly2009}
{Connolly}, J.~A.~D. 2009, Geochemistry, Geophysics, Geosystems, 10, Q10014, \dodoi{10.1029/2009GC002540}

\bibitem[{{Daspute} {et~al.}(2025){Daspute}, {Wandel}, {Kopparapu}, {Perdelwitz}, {Teklu}, \& {Tal-Or}}]{Daspute2025}
{Daspute}, M., {Wandel}, A., {Kopparapu}, R.~K., {et~al.} 2025, \apj, 979, 158, \dodoi{10.3847/1538-4357/ad9ba9}

\bibitem[{{de Wit} {et~al.}(2018){de Wit}, {Wakeford}, {Lewis}, {Delrez}, {Gillon}, {Selsis}, {Leconte}, {Demory}, {Bolmont}, {Bourrier}, {Burgasser}, {Grimm}, {Jehin}, {Lederer}, {Owen}, {Stamenkovi{\'c}}, \& {Triaud}}]{deWit2018}
{de Wit}, J., {Wakeford}, H.~R., {Lewis}, N.~K., {et~al.} 2018, Nature Astronomy, 2, 214, \dodoi{10.1038/s41550-017-0374-z}

\bibitem[{{Desai} {et~al.}(2024){Desai}, {Turtelboom}, {Harada}, {Dressing}, {Rice}, {Murphy}, {Brinkman}, {Chontos}, {Crossfield}, {Dai}, {Hill}, {Fetherolf}, {Giacalone}, {Howard}, {Huber}, {Isaacson}, {Kane}, {Lubin}, {MacDougall}, {Mayo}, {Mo{\v{c}}nik}, {Polanski}, {Rice}, {Robertson}, {Rubenzahl}, {Van Zandt}, {Weiss}, {Bieryla}, {Buchhave}, {Jenkins}, {Kostov}, {Levine}, {Lillo-Box}, {Paegert}, {Rabus}, {Seager}, {Stassun}, {Ting}, {Watanabe}, \& {Winn}}]{Desai2024}
{Desai}, A., {Turtelboom}, E.~V., {Harada}, C.~K., {et~al.} 2024, \aj, 167, 194, \dodoi{10.3847/1538-3881/ad29ee}

\bibitem[{{Dobos} {et~al.}(2019){Dobos}, {Barr}, \& {Kiss}}]{Dobos2019}
{Dobos}, V., {Barr}, A.~C., \& {Kiss}, L.~L. 2019, \aap, 624, A2, \dodoi{10.1051/0004-6361/201834254}

\bibitem[{{Dong} {et~al.}(2018){Dong}, {Jin}, {Lingam}, {Airapetian}, {Ma}, \& {van der Holst}}]{Dong2018}
{Dong}, C., {Jin}, M., {Lingam}, M., {et~al.} 2018, Proceedings of the National Academy of Science, 115, 260, \dodoi{10.1073/pnas.1708010115}

\bibitem[{{Dorn} {et~al.}(2017{\natexlab{a}}){Dorn}, {Hinkel}, \& {Venturini}}]{Dorn2017b}
{Dorn}, C., {Hinkel}, N.~R., \& {Venturini}, J. 2017{\natexlab{a}}, \aap, 597, A38, \dodoi{10.1051/0004-6361/201628749}

\bibitem[{{Dorn} {et~al.}(2015){Dorn}, {Khan}, {Heng}, {Connolly}, {Alibert}, {Benz}, \& {Tackley}}]{Dorn2015}
{Dorn}, C., {Khan}, A., {Heng}, K., {et~al.} 2015, \aap, 577, A83, \dodoi{10.1051/0004-6361/201424915}

\bibitem[{{Dorn} \& {Lichtenberg}(2021)}]{Dorn2021}
{Dorn}, C., \& {Lichtenberg}, T. 2021, \apjl, 922, L4, \dodoi{10.3847/2041-8213/ac33af}

\bibitem[{{Dorn} {et~al.}(2017{\natexlab{b}}){Dorn}, {Venturini}, {Khan}, {Heng}, {Alibert}, {Helled}, {Rivoldini}, \& {Benz}}]{Dorn2017}
{Dorn}, C., {Venturini}, J., {Khan}, A., {et~al.} 2017{\natexlab{b}}, \aap, 597, A37, \dodoi{10.1051/0004-6361/201628708}

\bibitem[{{Dorogokupets} {et~al.}(2017){Dorogokupets}, {Dymshits}, {Litasov}, \& {Sokolova}}]{Dorogokupets17}
{Dorogokupets}, P.~I., {Dymshits}, A.~M., {Litasov}, K.~D., \& {Sokolova}, T.~S. 2017, Scientific Reports, 7, 41863, \dodoi{10.1038/srep41863}

\bibitem[{{Dorogokupets} {et~al.}(2015){Dorogokupets}, {Dymshits}, {Sokolova}, {Danilov}, \& {Litasov}}]{Dorogokupets2015}
{Dorogokupets}, P.~I., {Dymshits}, A.~M., {Sokolova}, T.~S., {Danilov}, B.~S., \& {Litasov}, K.~D. 2015, Russian Geology and Geophysics, 56, 172, \dodoi{10.1016/j.rgg.2015.01.011}

\bibitem[{{Dunaeva} {et~al.}(2010){Dunaeva}, {Antsyshkin}, \& {Kuskov}}]{Dunaeva2010}
{Dunaeva}, A.~N., {Antsyshkin}, D.~V., \& {Kuskov}, O.~L. 2010, Solar System Research, 44, 202, \dodoi{10.1134/S0038094610030044}

\bibitem[{{Essack} {et~al.}(2023){Essack}, {Shporer}, {Burt}, {Seager}, {Cambioni}, {Lin}, {Collins}, {Mamajek}, {Stassun}, {Ricker}, {Vanderspek}, {Latham}, {Winn}, {Jenkins}, {Butler}, {Charbonneau}, {Collins}, {Crane}, {Gan}, {Hellier}, {Howell}, {Irwin}, {Mann}, {Ramadhan}, {Shectman}, {Teske}, {Yee}, {Mireles}, {Quintana}, {Tenenbaum}, {Torres}, \& {Furlan}}]{Essack2023}
{Essack}, Z., {Shporer}, A., {Burt}, J.~A., {et~al.} 2023, \aj, 165, 47, \dodoi{10.3847/1538-3881/ac9c5b}

\bibitem[{{Fauchez} {et~al.}(2019){Fauchez}, {Turbet}, {Villanueva}, {Wolf}, {Arney}, {Kopparapu}, {Lincowski}, {Mandell}, {de Wit}, {Pidhorodetska}, {Domagal-Goldman}, \& {Stevenson}}]{Fauchez2019}
{Fauchez}, T.~J., {Turbet}, M., {Villanueva}, G.~L., {et~al.} 2019, \apj, 887, 194, \dodoi{10.3847/1538-4357/ab5862}

\bibitem[{{Feistel} \& {Wagner}(2006)}]{Feistel2006}
{Feistel}, R., \& {Wagner}, W. 2006, Journal of Physical and Chemical Reference Data, 35, 1021, \dodoi{10.1063/1.2183324}

\bibitem[{{Gillon} {et~al.}(2016){Gillon}, {Jehin}, {Lederer}, {Delrez}, {de Wit}, {Burdanov}, {Van Grootel}, {Burgasser}, {Triaud}, {Opitom}, {Demory}, {Sahu}, {Bardalez Gagliuffi}, {Magain}, \& {Queloz}}]{Gillon2016}
{Gillon}, M., {Jehin}, E., {Lederer}, S.~M., {et~al.} 2016, \nat, 533, 221, \dodoi{10.1038/nature17448}

\bibitem[{{Gillon} {et~al.}(2017){Gillon}, {Triaud}, {Demory}, {Jehin}, {Agol}, {Deck}, {Lederer}, {de Wit}, {Burdanov}, {Ingalls}, {Bolmont}, {Leconte}, {Raymond}, {Selsis}, {Turbet}, {Barkaoui}, {Burgasser}, {Burleigh}, {Carey}, {Chaushev}, {Copperwheat}, {Delrez}, {Fernand es}, {Holdsworth}, {Kotze}, {Van Grootel}, {Almleaky}, {Benkhaldoun}, {Magain}, \& {Queloz}}]{trappist1first}
{Gillon}, M., {Triaud}, A. H.~M.~J., {Demory}, B.-O., {et~al.} 2017, \nat, 542, 456, \dodoi{10.1038/nature21360}

\bibitem[{Grande {et~al.}(2022)Grande, Pham, Smith, Boisvert, Huang, Smith, Goldman, Belof, Tschauner, Steffen, \& Salamat}]{Grande2022}
Grande, Z.~M., Pham, C.~H., Smith, D., {et~al.} 2022, Phys. Rev. B, 105, 104109, \dodoi{10.1103/PhysRevB.105.104109}

\bibitem[{{Grasset} {et~al.}(2009){Grasset}, {Schneider}, \& {Sotin}}]{Grasset2009}
{Grasset}, O., {Schneider}, J., \& {Sotin}, C. 2009, \apj, 693, 722, \dodoi{10.1088/0004-637X/693/1/722}

\bibitem[{Greene {et~al.}(2023)Greene, Bell, Ducrot, Dyrek, Lagage, \& Fortney}]{greene2023thermal}
Greene, T.~P., Bell, T.~J., Ducrot, E., {et~al.} 2023, Nature, 1

\bibitem[{{Grimm} {et~al.}(2018){Grimm}, {Demory}, {Gillon}, {Dorn}, {Agol}, {Burdanov}, {Delrez}, {Sestovic}, {Triaud}, {Turbet}, {Bolmont}, {Caldas}, {Wit}, {Jehin}, {Leconte}, {Raymond}, {Grootel}, {Burgasser}, {Carey}, {Fabrycky}, {Heng}, {Hernandez}, {Ingalls}, {Lederer}, {Selsis}, \& {Queloz}}]{TRAPPIST1}
{Grimm}, S.~L., {Demory}, B.-O., {Gillon}, M., {et~al.} 2018, \aap, 613, A68, \dodoi{10.1051/0004-6361/201732233}

\bibitem[{{Hakim} {et~al.}(2018){Hakim}, {Rivoldini}, {Van Hoolst}, {Cottenier}, {Jaeken}, {Chust}, \& {Steinle-Neumann}}]{Hakim2018}
{Hakim}, K., {Rivoldini}, A., {Van Hoolst}, T., {et~al.} 2018, \icarus, 313, 61, \dodoi{10.1016/j.icarus.2018.05.005}

\bibitem[{{Haldemann} {et~al.}(2020){Haldemann}, {Alibert}, {Mordasini}, \& {Benz}}]{Haldemann2020}
{Haldemann}, J., {Alibert}, Y., {Mordasini}, C., \& {Benz}, W. 2020, \aap, 643, A105, \dodoi{10.1051/0004-6361/202038367}

\bibitem[{{Haldemann} {et~al.}(2022){Haldemann}, {Ksoll}, {Walter}, {Alibert}, {Klessen}, {Benz}, {Koethe}, {Ardizzone}, \& {Rother}}]{Haldemann2022}
{Haldemann}, J., {Ksoll}, V., {Walter}, D., {et~al.} 2022, arXiv e-prints, arXiv:2202.00027, \dodoi{10.48550/arXiv.2202.00027}

\bibitem[{{Harper} {et~al.}(2015){Harper}, {Weinstein}, \& {Simon}}]{ternaryplot}
{Harper}, M., {Weinstein}, B., \& {Simon}, C. 2015, Zenodo 10.5281/zenodo.594435, \dodoi{10.5281/zenodo.594435}

\bibitem[{Harris {et~al.}(2020)Harris, Millman, van~der Walt, Gommers, Virtanen, Cournapeau, Wieser, Taylor, Berg, Smith, Kern, Picus, Hoyer, van Kerkwijk, Brett, Haldane, del R{\'{i}}o, Wiebe, Peterson, G{\'{e}}rard-Marchant, Sheppard, Reddy, Weckesser, Abbasi, Gohlke, \& Oliphant}]{numpy}
Harris, C.~R., Millman, K.~J., van~der Walt, S.~J., {et~al.} 2020, Nature, 585, 357, \dodoi{10.1038/s41586-020-2649-2}

\bibitem[{{Howard} {et~al.}(2025){Howard}, {Sinukoff}, {Blunt}, {Petigura}, {Crossfield}, {Isaacson}, {Kosiarek}, {Rubenzahl}, {Brewer}, {Fulton}, {Dressing}, {Hirsch}, {Knutson}, {Livingston}, {Mills}, {Roy}, {Weiss}, {Benneke}, {Ciardi}, {Christiansen}, {Cochran}, {Crepp}, {Gonzales}, {Hansen}, {Hardegree-Ullman}, {Howell}, {L{\'e}pine}, {Martinez}, {Rogers}, {Schlieder}, {Werner}, {Polanski}, {Angelo}, {Beard}, {Behmard}, {Bouma}, {Brinkman}, {Chontos}, {Dai}, {Dalba}, {Giacalone}, {Grunblatt}, {Hill}, {Kane}, {Lubin}, {Mayo}, {Mocnik}, {Akana Murphy}, {Rice}, {Rosenthal}, {Tyler}, {Van Zandt}, \& {Yee}}]{Howard2025}
{Howard}, A.~W., {Sinukoff}, E., {Blunt}, S., {et~al.} 2025, arXiv e-prints, arXiv:2502.04436.
\newblock \doarXiv{2502.04436}

\bibitem[{{Howe} {et~al.}(2014){Howe}, {Burrows}, \& {Verne}}]{Howe2014}
{Howe}, A.~R., {Burrows}, A., \& {Verne}, W. 2014, \apj, 787, 173, \dodoi{10.1088/0004-637X/787/2/173}

\bibitem[{{Huang} {et~al.}(2022){Huang}, {Rice}, \& {Steffen}}]{Magrathea}
{Huang}, C., {Rice}, D.~R., \& {Steffen}, J.~H. 2022, \mnras, 513, 5256, \dodoi{10.1093/mnras/stac1133}

\bibitem[{Hunter(2007)}]{matplotlib}
Hunter, J.~D. 2007, Computing in Science \& Engineering, 9, 90, \dodoi{10.1109/MCSE.2007.55}

\bibitem[{{Kellermann} {et~al.}(2018){Kellermann}, {Becker}, \& {Redmer}}]{Kellermann2018}
{Kellermann}, C., {Becker}, A., \& {Redmer}, R. 2018, \aap, 615, A39, \dodoi{10.1051/0004-6361/201731775}

\bibitem[{{Kemmer} {et~al.}(2022){Kemmer}, {Dreizler}, {Kossakowski}, {Stock}, {Quirrenbach}, {Caballero}, {Amado}, {Collins}, {Espinoza}, {Herrero}, {Jenkins}, {Latham}, {Lillo-Box}, {Narita}, {Pall{\'e}}, {Reiners}, {Ribas}, {Ricker}, {Rodr{\'\i}guez}, {Seager}, {Vanderspek}, {Wells}, {Winn}, {Aceituno}, {B{\'e}jar}, {Barclay}, {Bluhm}, {Chaturvedi}, {Cifuentes}, {Collins}, {Cort{\'e}s-Contreras}, {Demory}, {Fausnaugh}, {Fukui}, {G{\'o}mez Maqueo Chew}, {Galad{\'\i}-Enr{\'\i}quez}, {Gan}, {Gillon}, {Golovin}, {Hatzes}, {Henning}, {Huang}, {Jeffers}, {Kaminski}, {Kunimoto}, {K{\"u}rster}, {L{\'o}pez-Gonz{\'a}lez}, {Lafarga}, {Luque}, {McCormac}, {Molaverdikhani}, {Montes}, {Morales}, {Passegger}, {Reffert}, {Sabin}, {Sch{\"o}fer}, {Schanche}, {Schlecker}, {Schroffenegger}, {Schwarz}, {Schweitzer}, {Sota}, {Tenenbaum}, {Trifonov}, {Vanaverbeke}, \& {Zechmeister}}]{Kemmer2022}
{Kemmer}, J., {Dreizler}, S., {Kossakowski}, D., {et~al.} 2022, \aap, 659, A17, \dodoi{10.1051/0004-6361/202142653}

\bibitem[{{Kuskov} \& {Kronrod}(2001)}]{Kuskov2001}
{Kuskov}, O.~L., \& {Kronrod}, V.~A. 2001, \icarus, 151, 204, \dodoi{10.1006/icar.2001.6611}

\bibitem[{{Lay} {et~al.}(2008){Lay}, {Hernlund}, \& {Buffett}}]{Lay2008}
{Lay}, T., {Hernlund}, J., \& {Buffett}, B.~A. 2008, Nature Geoscience, 1, 25, \dodoi{10.1038/ngeo.2007.44}

\bibitem[{{Lodders} {et~al.}(2009){Lodders}, {Palme}, \& {Gail}}]{Lodders2009}
{Lodders}, K., {Palme}, H., \& {Gail}, H.~P. 2009, Landolt B\&ouml;rnstein, 4B, 712, \dodoi{10.1007/978-3-540-88055-4_34}

\bibitem[{{Luo} {et~al.}(2024){Luo}, {Dorn}, \& {Deng}}]{Luo2024}
{Luo}, H., {Dorn}, C., \& {Deng}, J. 2024, Nature Astronomy, 8, 1399, \dodoi{10.1038/s41550-024-02347-z}

\bibitem[{{Luque} \& {Pall{\'e}}(2022)}]{Luque2022}
{Luque}, R., \& {Pall{\'e}}, E. 2022, Science, 377, 1211, \dodoi{10.1126/science.abl7164}

\bibitem[{{Lustig-Yaeger} {et~al.}(2019){Lustig-Yaeger}, {Meadows}, \& {Lincowski}}]{Lustig-Yaeger2019}
{Lustig-Yaeger}, J., {Meadows}, V.~S., \& {Lincowski}, A.~P. 2019, \aj, 158, 27, \dodoi{10.3847/1538-3881/ab21e0}

\bibitem[{{Madhusudhan} {et~al.}(2012){Madhusudhan}, {Lee}, \& {Mousis}}]{Madhusudhan2012}
{Madhusudhan}, N., {Lee}, K. K.~M., \& {Mousis}, O. 2012, \apjl, 759, L40, \dodoi{10.1088/2041-8205/759/2/L40}

\bibitem[{{Madhusudhan} {et~al.}(2021){Madhusudhan}, {Piette}, \& {Constantinou}}]{Madhusudhan2021}
{Madhusudhan}, N., {Piette}, A. A.~A., \& {Constantinou}, S. 2021, \apj, 918, 1, \dodoi{10.3847/1538-4357/abfd9c}

\bibitem[{{Mikal-Evans}(2022)}]{Mikal-Evans2022}
{Mikal-Evans}, T. 2022, \mnras, 510, 980, \dodoi{10.1093/mnras/stab3383}

\bibitem[{{Neil} \& {Rogers}(2020)}]{Neil2020}
{Neil}, A.~R., \& {Rogers}, L.~A. 2020, \apj, 891, 12, \dodoi{10.3847/1538-4357/ab6a92}

\bibitem[{{Nixon} \& {Madhusudhan}(2021)}]{Nixon2021}
{Nixon}, M.~C., \& {Madhusudhan}, N. 2021, \mnras, 505, 3414, \dodoi{10.1093/mnras/stab1500}

\bibitem[{{Noack} \& {Lasbleis}(2020)}]{Noack2020}
{Noack}, L., \& {Lasbleis}, M. 2020, \aap, 638, A129, \dodoi{10.1051/0004-6361/202037723}

\bibitem[{Nomura {et~al.}(2014)Nomura, Hirose, Uesugi, Ohishi, Tsuchiyama, Miyake, \& Ueno}]{Nomura2014}
Nomura, R., Hirose, K., Uesugi, K., {et~al.} 2014, Science, 343, 522, \dodoi{10.1126/science.1248186}

\bibitem[{{Oganov} \& {Ono}(2004)}]{Oganov04}
{Oganov}, A.~R., \& {Ono}, S. 2004, \nat, 430, 445, \dodoi{10.1038/nature02701}

\bibitem[{{Ono} \& {Oganov}(2005)}]{Ono05}
{Ono}, S., \& {Oganov}, A.~R. 2005, Earth and Planetary Science Letters, 236, 914, \dodoi{10.1016/j.epsl.2005.06.001}

\bibitem[{{Padovan} {et~al.}(2018){Padovan}, {Spohn}, {Baumeister}, {Tosi}, {Breuer}, {Csizmadia}, {Hellard}, \& {Sohl}}]{Padovan2018}
{Padovan}, S., {Spohn}, T., {Baumeister}, P., {et~al.} 2018, \aap, 620, A178, \dodoi{10.1051/0004-6361/201834181}

\bibitem[{Papakonstantinou \& Tapia(2013)}]{secant}
Papakonstantinou, J.~M., \& Tapia, R.~A. 2013, The American Mathematical Monthly, 120, pp. 500.
\newblock \url{https://www.jstor.org/stable/10.4169/amer.math.monthly.120.06.500}

\bibitem[{{Piaulet} {et~al.}(2023){Piaulet}, {Benneke}, {Almenara}, {Dragomir}, {Knutson}, {Thorngren}, {Peterson}, {Crossfield}, {Kempton}, {Kubyshkina}, {Howard}, {Angus}, {Isaacson}, {Weiss}, {Beichman}, {Fortney}, {Fossati}, {Lammer}, {McCullough}, {Morley}, \& {Wong}}]{Piaulet2023}
{Piaulet}, C., {Benneke}, B., {Almenara}, J.~M., {et~al.} 2023, Nature Astronomy, 7, 206, \dodoi{10.1038/s41550-022-01835-4}

\bibitem[{{Plotnykov} \& {Valencia}(2020)}]{Plotnykov2020}
{Plotnykov}, M., \& {Valencia}, D. 2020, \mnras, 499, 932, \dodoi{10.1093/mnras/staa2615}

\bibitem[{{Pozzo} {et~al.}(2012){Pozzo}, {Davies}, {Gubbins}, \& {Alf{\`e}}}]{2012Natur.485..355P}
{Pozzo}, M., {Davies}, C., {Gubbins}, D., \& {Alf{\`e}}, D. 2012, \nat, 485, 355, \dodoi{10.1038/nature11031}

\bibitem[{{Putirka} \& {Rarick}(2019)}]{Putirka2019}
{Putirka}, K.~D., \& {Rarick}, J.~C. 2019, American Mineralogist, 104, 817, \dodoi{10.2138/am-2019-6787}

\bibitem[{{Rauer} {et~al.}(2024){Rauer}, {Aerts}, {Cabrera}, {Deleuil}, {Erikson}, {Gizon}, {Goupil}, {Heras}, {Lorenzo-Alvarez}, {Marliani}, {Martin-Garcia}, {Mas-Hesse}, {O'Rourke}, {Osborn}, {Pagano}, {Piotto}, {Pollacco}, {Ragazzoni}, {Ramsay}, {Udry}, {Appourchaux}, {Benz}, {Brandeker}, {G{\"u}del}, {Janot-Pacheco}, {Kabath}, {Kjeldsen}, {Min}, {Santos}, {Smith}, {Suarez}, {Werner}, {Aboudan}, {Abreu}, {Acu a}, {Adams}, {Adibekyan}, {Affer}, {Agneray}, {Agnor}, {Aguirre B{\o}rsen-Koch}, {Ahmed}, {Aigrain}, {Al-Bahlawan}, {Alcacera Gil}, {Alei}, {Alencar}, {Alexander}, {Alfonso-Garz{\'o}n}, {Alibert}, {Allende Prieto}, {Almeida}, {Alonso Sobrino}, {Altavilla}, {Althaus}, {Alonso Alvarez Trujillo}, {Amarsi}, {Ammler-von Eiff}, {Am{\^o}res}, {Andrade}, {Antoniadis-Karnavas}, {Ant{\'o}nio}, {Aparicio del Moral}, {Appolloni}, {Arena}, {Armstrong}, {Aroca Aliaga}, {Asplund}, {Audenaert}, {Auricchio}, {Avelino}, {Baeke}, {Bailli{\'e}}, {Balado}, {Ballber Balaguer{\'o}}, {Balestra}, {Ball}, {Ballans}, {Ballot},
  {Barban}, {Barbary}, {Barbieri}, {Barcel{\'o} Forteza}, {Barker}, {Barklem}, {Barnes}, {Barrado Navascues}, {Barragan}, {Baruteau}, {Basu}, {Baudin}, {Baumeister}, {Bayliss}, {Bazot}, {Beck}, {Bedding}, {Belkacem}, {Bellinger}, {Benatti}, {Benomar}, {B{\'e}rard}, {Bergemann}, {Bergomi}, {Bernardo}, {Biazzo}, {Bignamini}, {Bigot}, {Billot}, {Binet}, {Biondi}, {Biondi}, {Birch}, {Bitsch}, {Bluhm Ceballos}, {B{\'o}di}, {Bogn{\'a}r}, {Boisse}, {Bolmont}, {Bonanno}, {Bonavita}, {Bonfanti}, {Bonfils}, {Bonito}, {Bonomo}, {B{\"o}rner}, {Boro Saikia}, {Borreguero Mart{\'\i}n}, {Borsa}, {Borsato}, {Bossini}, {Bouchy}, {Bou{\'e}}, {Boufleur}, {Boumier}, {Bourrier}, {Bowman}, {Bozzo}, {Bradley}, {Bray}, {Bressan}, {Breton}, {Brienza}, {Brito}, {Brogi}, {Brown}, {Brown}, {Brun}, {Bruno}, {Bruns}, {Buchhave}, {Bugnet}, {Buldgen}, {Burgess}, {Busatta}, {Busso}, {Buzasi}, {Caballero}, {Cabral}, {Cabrero Gomez}, {Calderone}, {Cameron}, {Cameron}, {Campante}, {Campos Gestal}, {Canto Martins}, {Cara}, {Carone}, {Carrasco},
  {Casagrande}, {Casewell}, {Cassisi}, {Castellani}, {Castro}, {Catala}, {Catal{\'a}n Fern{\'a}ndez}, {Catelan}, {Cegla}, {Cerruti}, {Cessa}, {Chadid}, {Chaplin}, {Charpinet}, {Chiappini}, {Chiarucci}, {Chiavassa}, {Chinellato}, {Chirulli}, {Christensen-Dalsgaard}, {Church}, {Claret}, {Clarke}, {Claudi}, {Clermont}, {Coelho}, {Coelho}, {Cogato}, {Colom{\'e}}, {Condamin}, {Conde Garc{\'\i}a}, \& {Conseil}}]{Rauer2024}
{Rauer}, H., {Aerts}, C., {Cabrera}, J., {et~al.} 2024, arXiv e-prints, arXiv:2406.05447, \dodoi{10.48550/arXiv.2406.05447}

\bibitem[{{Rigby} {et~al.}(2024){Rigby}, {Pica-Ciamarra}, {Holmberg}, {Madhusudhan}, {Constantinou}, {Schaefer}, {Deng}, {Lee}, \& {Moses}}]{Rigby2024}
{Rigby}, F.~E., {Pica-Ciamarra}, L., {Holmberg}, M., {et~al.} 2024, \apj, 975, 101, \dodoi{10.3847/1538-4357/ad6c38}

\bibitem[{{Rodr{\'\i}guez Mart{\'\i}nez} {et~al.}(2021){Rodr{\'\i}guez Mart{\'\i}nez}, {Stevens}, {Gaudi}, {Schulze}, {Panero}, {Johnson}, \& {Wang}}]{RodriguezMartinez2021}
{Rodr{\'\i}guez Mart{\'\i}nez}, R., {Stevens}, D.~J., {Gaudi}, B.~S., {et~al.} 2021, \apj, 911, 84, \dodoi{10.3847/1538-4357/abe941}

\bibitem[{{Rogers} \& {Seager}(2010)}]{Rogers2010}
{Rogers}, L.~A., \& {Seager}, S. 2010, \apj, 712, 974, \dodoi{10.1088/0004-637X/712/2/974}

\bibitem[{{Sakai} {et~al.}(2016){Sakai}, {Dekura}, \& {Hirao}}]{Sakai16}
{Sakai}, T., {Dekura}, H., \& {Hirao}, N. 2016, Scientific Reports, 6, 22652, \dodoi{10.1038/srep22652}

\bibitem[{{Santos} {et~al.}(2015){Santos}, {Adibekyan}, {Mordasini}, {Benz}, {Delgado-Mena}, {Dorn}, {Buchhave}, {Figueira}, {Mortier}, {Pepe}, {Santerne}, {Sousa}, \& {Udry}}]{Santos2015}
{Santos}, N.~C., {Adibekyan}, V., {Mordasini}, C., {et~al.} 2015, \aap, 580, L13, \dodoi{10.1051/0004-6361/201526850}

\bibitem[{{Schubert} {et~al.}(2007){Schubert}, {Anderson}, {Spohn}, \& {McKinnon}}]{Schubert2007}
{Schubert}, G., {Anderson}, J.~D., {Spohn}, T., \& {McKinnon}, W.~B. 2007, in Jupiter, 281

\bibitem[{{Seager} {et~al.}(2007){Seager}, {Kuchner}, {Hier-Majumder}, \& {Militzer}}]{Seager2007}
{Seager}, S., {Kuchner}, M., {Hier-Majumder}, C.~A., \& {Militzer}, B. 2007, \apj, 669, 1279, \dodoi{10.1086/521346}

\bibitem[{{Serrano} {et~al.}(2022){Serrano}, {Gandolfi}, {Mustill}, {Barrag{\'a}n}, {Korth}, {Dai}, {Redfield}, {Fridlund}, {Lam}, {D{\'\i}az}, {Grziwa}, {Collins}, {Livingston}, {Cochran}, {Hellier}, {Bellomo}, {Trifonov}, {Rodler}, {Alarcon}, {Jenkins}, {Latham}, {Ricker}, {Seager}, {Vanderspeck}, {Winn}, {Albrecht}, {Collins}, {Csizmadia}, {Daylan}, {Deeg}, {Esposito}, {Fausnaugh}, {Georgieva}, {Goffo}, {Guenther}, {Hatzes}, {Howell}, {Jensen}, {Luque}, {Mann}, {Murgas}, {Osborne}, {Palle}, {Persson}, {Rowden}, {Rudat}, {Smith}, {Twicken}, {Van Eylen}, \& {Ziegler}}]{Serrano2022}
{Serrano}, L.~M., {Gandolfi}, D., {Mustill}, A.~J., {et~al.} 2022, Nature Astronomy, 6, 736, \dodoi{10.1038/s41550-022-01641-y}

\bibitem[{{Shah} {et~al.}(2021){Shah}, {Alibert}, {Helled}, \& {Mezger}}]{Shah2021}
{Shah}, O., {Alibert}, Y., {Helled}, R., \& {Mezger}, K. 2021, \aap, 646, A162, \dodoi{10.1051/0004-6361/202038839}

\bibitem[{{Shim} \& {Duffy}(2000)}]{Shim2000}
{Shim}, S.-H., \& {Duffy}, T.~S. 2000, American Mineralogist, 85, 354, \dodoi{10.2138/am-2000-2-314}

\bibitem[{{Shorttle} {et~al.}(2024){Shorttle}, {Jordan}, {Nicholls}, {Lichtenberg}, \& {Bower}}]{Shorttle2024}
{Shorttle}, O., {Jordan}, S., {Nicholls}, H., {Lichtenberg}, T., \& {Bower}, D.~J. 2024, \apjl, 962, L8, \dodoi{10.3847/2041-8213/ad206e}

\bibitem[{{Smith} {et~al.}(2018){Smith}, {Fratanduono}, {Braun}, {Duffy}, {Wicks}, {Celliers}, {Ali}, {Fernandez-Pa{\~n}ella}, {Kraus}, {Swift}, {Collins}, \& {Eggert}}]{Smith2018}
{Smith}, R.~F., {Fratanduono}, D.~E., {Braun}, D.~G., {et~al.} 2018, Nature Astronomy, 2, 452, \dodoi{10.1038/s41550-018-0437-9}

\bibitem[{{Sohl} {et~al.}(2002){Sohl}, {Spohn}, {Breuer}, \& {Nagel}}]{Sohl2002}
{Sohl}, F., {Spohn}, T., {Breuer}, D., \& {Nagel}, K. 2002, \icarus, 157, 104, \dodoi{10.1006/icar.2002.6828}

\bibitem[{{Sotin} {et~al.}(2007){Sotin}, {Grasset}, \& {Mocquet}}]{Sotin2007}
{Sotin}, C., {Grasset}, O., \& {Mocquet}, A. 2007, Icarus, 191, 337, \dodoi{10.1016/j.icarus.2007.04.006}

\bibitem[{{Suissa} {et~al.}(2018){Suissa}, {Chen}, \& {Kipping}}]{Suissa2018}
{Suissa}, G., {Chen}, J., \& {Kipping}, D. 2018, \mnras, 476, 2613, \dodoi{10.1093/mnras/sty381}

\bibitem[{{Thomas} \& {Madhusudhan}(2016)}]{Thomas2016}
{Thomas}, S.~W., \& {Madhusudhan}, N. 2016, \mnras, 458, 1330, \dodoi{10.1093/mnras/stw321}

\bibitem[{{Tinetti} {et~al.}(2018){Tinetti}, {Drossart}, {Eccleston}, {Hartogh}, {Heske}, {Leconte}, {Micela}, {Ollivier}, {Pilbratt}, {Puig}, {Turrini}, {Vandenbussche}, {Wolkenberg}, {Beaulieu}, {Buchave}, {Ferus}, {Griffin}, {Guedel}, {Justtanont}, {Lagage}, {Machado}, {Malaguti}, {Min}, {N{\o}rgaard-Nielsen}, {Rataj}, {Ray}, {Ribas}, {Swain}, {Szabo}, {Werner}, {Barstow}, {Burleigh}, {Cho}, {Coud{\'e} du Foresto}, {Coustenis}, {Decin}, {Encrenaz}, {Galand}, {Gillon}, {Helled}, {Morales}, {Garc{\'\i}a Mu{\~n}oz}, {Moneti}, {Pagano}, {Pascale}, {Piccioni}, {Pinfield}, {Sarkar}, {Selsis}, {Tennyson}, {Triaud}, {Venot}, {Waldmann}, {Waltham}, {Wright}, {Amiaux}, {Augu{\`e}res}, {Berth{\'e}}, {Bezawada}, {Bishop}, {Bowles}, {Coffey}, {Colom{\'e}}, {Crook}, {Crouzet}, {Da Peppo}, {Sanz}, {Focardi}, {Frericks}, {Hunt}, {Kohley}, {Middleton}, {Morgante}, {Ottensamer}, {Pace}, {Pearson}, {Stamper}, {Symonds}, {Rengel}, {Renotte}, {Ade}, {Affer}, {Alard}, {Allard}, {Altieri}, {Andr{\'e}}, {Arena}, {Argyriou},
  {Aylward}, {Baccani}, {Bakos}, {Banaszkiewicz}, {Barlow}, {Batista}, {Bellucci}, {Benatti}, {Bernardi}, {B{\'e}zard}, {Blecka}, {Bolmont}, {Bonfond}, {Bonito}, {Bonomo}, {Brucato}, {Brun}, {Bryson}, {Bujwan}, {Casewell}, {Charnay}, {Pestellini}, {Chen}, {Ciaravella}, {Claudi}, {Cl{\'e}dassou}, {Damasso}, {Damiano}, {Danielski}, {Deroo}, {Di Giorgio}, {Dominik}, {Doublier}, {Doyle}, {Doyon}, {Drummond}, {Duong}, {Eales}, {Edwards}, {Farina}, {Flaccomio}, {Fletcher}, {Forget}, {Fossey}, {Fr{\"a}nz}, {Fujii}, {Garc{\'\i}a-Piquer}, {Gear}, {Geoffray}, {G{\'e}rard}, {Gesa}, {Gomez}, {Graczyk}, {Griffith}, {Grodent}, {Guarcello}, {Gustin}, {Hamano}, {Hargrave}, {Hello}, {Heng}, {Herrero}, {Hornstrup}, {Hubert}, {Ida}, {Ikoma}, {Iro}, {Irwin}, {Jarchow}, {Jaubert}, {Jones}, {Julien}, {Kameda}, {Kerschbaum}, {Kervella}, {Koskinen}, {Krijger}, {Krupp}, {Lafarga}, {Landini}, {Lellouch}, {Leto}, {Luntzer}, {Rank-L{\"u}ftinger}, {Maggio}, {Maldonado}, {Maillard}, {Mall}, {Marquette}, {Mathis}, {Maxted}, {Matsuo},
  {Medvedev}, {Miguel}, {Minier}, {Morello}, {Mura}, {Narita}, {Nascimbeni}, {Nguyen Tong}, {Noce}, {Oliva}, {Palle}, {Palmer}, {Pancrazzi}, {Papageorgiou}, {Parmentier}, {Perger}, {Petralia}, {Pezzuto}, {Pierrehumbert}, \& {Pillitteri}}]{Tinetti2018}
{Tinetti}, G., {Drossart}, P., {Eccleston}, P., {et~al.} 2018, Experimental Astronomy, 46, 135, \dodoi{10.1007/s10686-018-9598-x}

\bibitem[{{TRAPPIST-1 JWST Community Initiative} {et~al.}(2024){TRAPPIST-1 JWST Community Initiative}, {de Wit}, {Doyon}, {Rackham}, {Lim}, {Ducrot}, {Kreidberg}, {Benneke}, {Ribas}, {Berardo}, {Niraula}, {Iyer}, {Shapiro}, {Kostogryz}, {Witzke}, {Gillon}, {Agol}, {Meadows}, {Burgasser}, {Owen}, {Fortney}, {Selsis}, {Bello-Arufe}, {de Beurs}, {Bolmont}, {Cowan}, {Dong}, {Drake}, {Garcia}, {Greene}, {Haworth}, {Hu}, {Kane}, {Kervella}, {Koll}, {Krissansen-Totton}, {Lagage}, {Lichtenberg}, {Lustig-Yaeger}, {Lingam}, {Turbet}, {Seager}, {Barkaoui}, {Bell}, {Burdanov}, {Cadieux}, {Charnay}, {Cloutier}, {Cook}, {Correia}, {Dang}, {Daylan}, {Delrez}, {Edwards}, {Fauchez}, {Flagg}, {Fraschetti}, {Haqq-Misra}, {Huang}, {Iro}, {Jayawardhana}, {Jehin}, {Jin}, {Kite}, {Kitzmann}, {Kral}, {Lafreni{\`e}re}, {Libert}, {Liu}, {Mohanty}, {Morris}, {Murray}, {Piaulet}, {Pozuelos}, {Radica}, {Ranjan}, {Rathcke}, {Roy}, {Schwieterman}, {Turner}, {Triaud}, \& {Way}}]{TRAPPIST-1JWSTCommunityInitiative2024}
{TRAPPIST-1 JWST Community Initiative}, {de Wit}, J., {Doyon}, R., {et~al.} 2024, Nature Astronomy, 8, 810, \dodoi{10.1038/s41550-024-02298-5}

\bibitem[{{Tsai} {et~al.}(2021){Tsai}, {Innes}, {Lichtenberg}, {Taylor}, {Malik}, {Chubb}, \& {Pierrehumbert}}]{Tsai2021}
{Tsai}, S.-M., {Innes}, H., {Lichtenberg}, T., {et~al.} 2021, \apjl, 922, L27, \dodoi{10.3847/2041-8213/ac399a}

\bibitem[{{Turbet} {et~al.}(2020){Turbet}, {Bolmont}, {Ehrenreich}, {Gratier}, {Leconte}, {Selsis}, {Hara}, \& {Lovis}}]{Turbet2020}
{Turbet}, M., {Bolmont}, E., {Ehrenreich}, D., {et~al.} 2020, \aap, 638, A41, \dodoi{10.1051/0004-6361/201937151}

\bibitem[{{Unterborn} {et~al.}(2023){Unterborn}, {Desch}, {Haldemann}, {Lorenzo}, {Schulze}, {Hinkel}, \& {Panero}}]{Unterborn2023}
{Unterborn}, C.~T., {Desch}, S.~J., {Haldemann}, J., {et~al.} 2023, \apj, 944, 42, \dodoi{10.3847/1538-4357/acaa3b}

\bibitem[{{Unterborn} {et~al.}(2018){Unterborn}, {Desch}, {Hinkel}, \& {Lorenzo}}]{Unterborn2018}
{Unterborn}, C.~T., {Desch}, S.~J., {Hinkel}, N.~R., \& {Lorenzo}, A. 2018, Nature Astronomy, 2, 297, \dodoi{10.1038/s41550-018-0411-6}

\bibitem[{{Unterborn} \& {Panero}(2019)}]{Unterborn2019}
{Unterborn}, C.~T., \& {Panero}, W.~R. 2019, Journal of Geophysical Research (Planets), 124, 1704, \dodoi{10.1029/2018JE005844}

\bibitem[{{Valencia} {et~al.}(2007{\natexlab{a}}){Valencia}, {Sasselov}, \& {O'Connell}}]{Valencia2007}
{Valencia}, D., {Sasselov}, D.~D., \& {O'Connell}, R.~J. 2007{\natexlab{a}}, \apj, 656, 545, \dodoi{10.1086/509800}

\bibitem[{{Valencia} {et~al.}(2007{\natexlab{b}}){Valencia}, {Sasselov}, \& {O'Connell}}]{Valencia2007tern}
---. 2007{\natexlab{b}}, \apj, 665, 1413, \dodoi{10.1086/519554}

\bibitem[{{Van der Velden}(2020)}]{cmasher}
{Van der Velden}, E. 2020, The Journal of Open Source Software, 5, 2004, \dodoi{10.21105/joss.02004}

\bibitem[{{Vazan} {et~al.}(2013){Vazan}, {Kovetz}, {Podolak}, \& {Helled}}]{Vazan2013}
{Vazan}, A., {Kovetz}, A., {Podolak}, M., \& {Helled}, R. 2013, \mnras, 434, 3283, \dodoi{10.1093/mnras/stt1248}

\bibitem[{{Vazan} {et~al.}(2022){Vazan}, {Sari}, \& {Kessel}}]{Vazan2022}
{Vazan}, A., {Sari}, R., \& {Kessel}, R. 2022, \apj, 926, 150, \dodoi{10.3847/1538-4357/ac458c}

\bibitem[{{Vida} \& {Roettenbacher}(2018)}]{Vida2018}
{Vida}, K., \& {Roettenbacher}, R.~M. 2018, \aap, 616, A163, \dodoi{10.1051/0004-6361/201833194}

\bibitem[{{Weeks} {et~al.}(2024){Weeks}, {Van Eylen}, {Huber}, {Kawata}, {Stokholm}, {Aguirre B{\o}rsen-Koch}, {Pinilla}, {Lysgaard R{\o}rsted}, {Lykke Winther}, \& {Berger}}]{Weeks2024}
{Weeks}, A., {Van Eylen}, V., {Huber}, D., {et~al.} 2024, arXiv e-prints, arXiv:2411.17358, \dodoi{10.48550/arXiv.2411.17358}

\bibitem[{{Wicks} {et~al.}(2018){Wicks}, {Smith}, {Fratanduono}, {Coppari}, {Kraus}, {Newman}, {Rygg}, {Eggert}, \& {Duffy}}]{Wicks2018}
{Wicks}, J.~K., {Smith}, R.~F., {Fratanduono}, D.~E., {et~al.} 2018, Science Advances, 4, eaao5864, \dodoi{10.1126/sciadv.aao5864}

\bibitem[{{Winters} {et~al.}(2022){Winters}, {Cloutier}, {Medina}, {Irwin}, {Charbonneau}, {Astudillo-Defru}, {Bonfils}, {Howard}, {Isaacson}, {Bean}, {Seifahrt}, {Teske}, {Eastman}, {Twicken}, {Collins}, {Jensen}, {Quinn}, {Payne}, {Kristiansen}, {Spencer}, {Vanderburg}, {Zechmeister}, {Weiss}, {Wang}, {Wang}, {Udry}, {Terentev}, {St{\"u}rmer}, {Stef{\'a}nsson}, {Shporer}, {Shectman}, {Sefako}, {Schwengeler}, {Schwarz}, {Scarsdale}, {Rubenzahl}, {Roy}, {Rosenthal}, {Robertson}, {Petigura}, {Pepe}, {Omohundro}, {Murphy}, {Murgas}, {Mo{\v{c}}nik}, {Montet}, {Mennickent}, {Mayo}, {Massey}, {Lubin}, {Lovis}, {Lewin}, {Kasper}, {Kane}, {Jenkins}, {Huber}, {Horne}, {Hill}, {Gorrini}, {Giacalone}, {Fulton}, {Forveille}, {Figueira}, {Fetherolf}, {Dressing}, {D{\'\i}az}, {Delfosse}, {Dalba}, {Dai}, {Cort{\'e}s}, {Crossfield}, {Crane}, {Conti}, {Collins}, {Chontos}, {Butler}, {Brown}, {Brady}, {Behmard}, {Beard}, {Batalha}, \& {Almenara}}]{Winters2022}
{Winters}, J.~G., {Cloutier}, R., {Medina}, A.~A., {et~al.} 2022, \aj, 163, 168, \dodoi{10.3847/1538-3881/ac50a9}

\bibitem[{{Yang} {et~al.}(2024){Yang}, {Song}, {Wu}, {Mao}, \& {Zhang}}]{Yang2024}
{Yang}, Z., {Song}, Z., {Wu}, Z., {Mao}, H.-k., \& {Zhang}, L. 2024, Proceedings of the National Academy of Science, 121, e2401281121, \dodoi{10.1073/pnas.2401281121}

\bibitem[{Yin {et~al.}(2022)Yin, Zhang, Zhang, Zhai, \& Liu}]{yin2022electrical}
Yin, Y., Zhang, Q., Zhang, Y., Zhai, S., \& Liu, Y. 2022, Acta Geochimica, 41, 665

\bibitem[{{Zeng} {et~al.}(2016){Zeng}, {Sasselov}, \& {Jacobsen}}]{Zeng2016}
{Zeng}, L., {Sasselov}, D.~D., \& {Jacobsen}, S.~B. 2016, \apj, 819, 127, \dodoi{10.3847/0004-637X/819/2/127}

\bibitem[{{Zeng} \& {Seager}(2008)}]{Zeng2008}
{Zeng}, L., \& {Seager}, S. 2008, \pasp, 120, 983, \dodoi{10.1086/591807}

\bibitem[{{Zeng} {et~al.}(2019){Zeng}, {Jacobsen}, {Sasselov}, {Petaev}, {Vanderburg}, {Lopez-Morales}, {Perez-Mercader}, {Mattsson}, {Li}, {Heising}, {Bonomo}, {Damasso}, {Berger}, {Cao}, {Levi}, \& {Wordsworth}}]{Zeng2019}
{Zeng}, L., {Jacobsen}, S.~B., {Sasselov}, D.~D., {et~al.} 2019, Proceedings of the National Academy of Science, 116, 9723, \dodoi{10.1073/pnas.1812905116}

\bibitem[{{Zhao} {et~al.}(2023){Zhao}, {Ni}, \& {Liu}}]{2023ApJS..269....1Z}
{Zhao}, Y., {Ni}, D., \& {Liu}, Z. 2023, \apjs, 269, 1, \dodoi{10.3847/1538-4365/acf31a}

\end{thebibliography}
\bibliographystyle{aasjournal}



\end{document}